\documentclass[fleqn,usenatbib]{mnras}

\usepackage{newtxtext,newtxmath}

\usepackage[T1]{fontenc}
\usepackage{ae,aecompl}

\usepackage{graphicx}	
\usepackage{amsmath}	
\usepackage{amssymb}	
\usepackage{wasysym}
\usepackage{url}
\usepackage{pdflscape}
\usepackage{afterpage}
\usepackage{epstopdf}
\usepackage{deluxetable}

\title[Co-spatial UV-optical STIS Spectra of Six PNe]{Co-spatial UV-optical HST/STIS Spectra of Six Planetary Nebulae: Nebular and Stellar Properties$^{1}$}

\author[Miller et al.]{
Timothy R. Miller,$^{2,9}$\thanks{E-mail: tim@huskers.unl.edu}
Richard B. C. Henry,$^{2}$
Bruce Balick,$^{3}$
Karen B. Kwitter,$^{4}$\newauthor
Reginald J. Dufour,$^{5}$
Richard A. Shaw,$^{6}$
and Romano L. M. Corradi$^{7,8}$
\\
$^{1}$Based on observations with the NASA/ESA \textit{Hubble Space Telescope} obtained at the Space Telescope Science Institute, \\which is operated by the Association of Universities for Research in Astronomy, Incorporated, under NASA contract NAS5-26555.\\
$^{2}$Department of Physics and Astronomy, University of Oklahoma, Norman, OK 73019\\
$^{3}$Department of Astronomy, University of Washington, Seattle, WA 98195\\
$^{4}$Department of Astronomy, Williams College, Williamstown, MA 01267\\
$^{5}$Department of Physics and Astronomy, Rice University, Houston, TX 77251\\
$^{6}$Space Telescope Science Institute, Baltimore, MD 21218\\
$^{7}$GRANTECAN, Cuesta de San Jos\'{e} s/n, E-38712, Bre\~{n}a Baja, La Palma, Spain\\
$^{8}$Instituto de Astrof\'{i}sica de Canarias, V\'{i}a L\'{a}ctea s/n, E-38200, La Laguna, Tenerife, Spain\\
$^{9}$Department of Physics, Virginia Tech, Blacksburg, VA 24061
}

\date{Accepted XXX. Received YYY; in original form ZZZ}

\pubyear{2017}

\begin{document}
\label{firstpage}
\pagerange{\pageref{firstpage}--\pageref{lastpage}}
\maketitle

\begin{abstract}
This paper represents the conclusion of a project that had two main goals: (1) to investigate to what extent planetary nebulae (PNe) are chemically homogeneous; and (2) to provide physical constraints on the central star properties of each PN. We accomplished the first goal by using HST/STIS spectra to measure the abundances of seven elements in numerous spatial regions within each of six PN (IC 2165, IC 3568, NGC 2440, NGC 5315, NGC 5882, and NGC 7662). The second goal was achieved by computing a photoionization model of each nebula, using our observed emission line strengths as constraints. The major finding of our study is that the nebular abundances of He, C, N, O, Ne, S, and Ar are consistent with a chemically homogeneous picture for each PN. Additionally, we found through experimenting with three different density profiles (constant, Gaussian, and Gaussian with a power-law) that the determination of the central star's temperature and luminosity is only slightly sensitive to the profile choice. Lastly, post-AGB evolutionary model predictions of temperature and luminosity available in the literature were plotted along with the values inferred from the photoionization model analysis to yield initial and final mass estimates of each central star.
\end{abstract}




\section{Introduction}

\begin{figure*}
\caption{The locations of the extracted spectra where all seven gratings spatially overlap one another from STIS. In addition to the individual regions, a Full region for each object was extracted over all the regions shown. The double slits in NGC 2440 are the result of a pointing error discussed in the text.\newline} 
\includegraphics[scale=0.255]{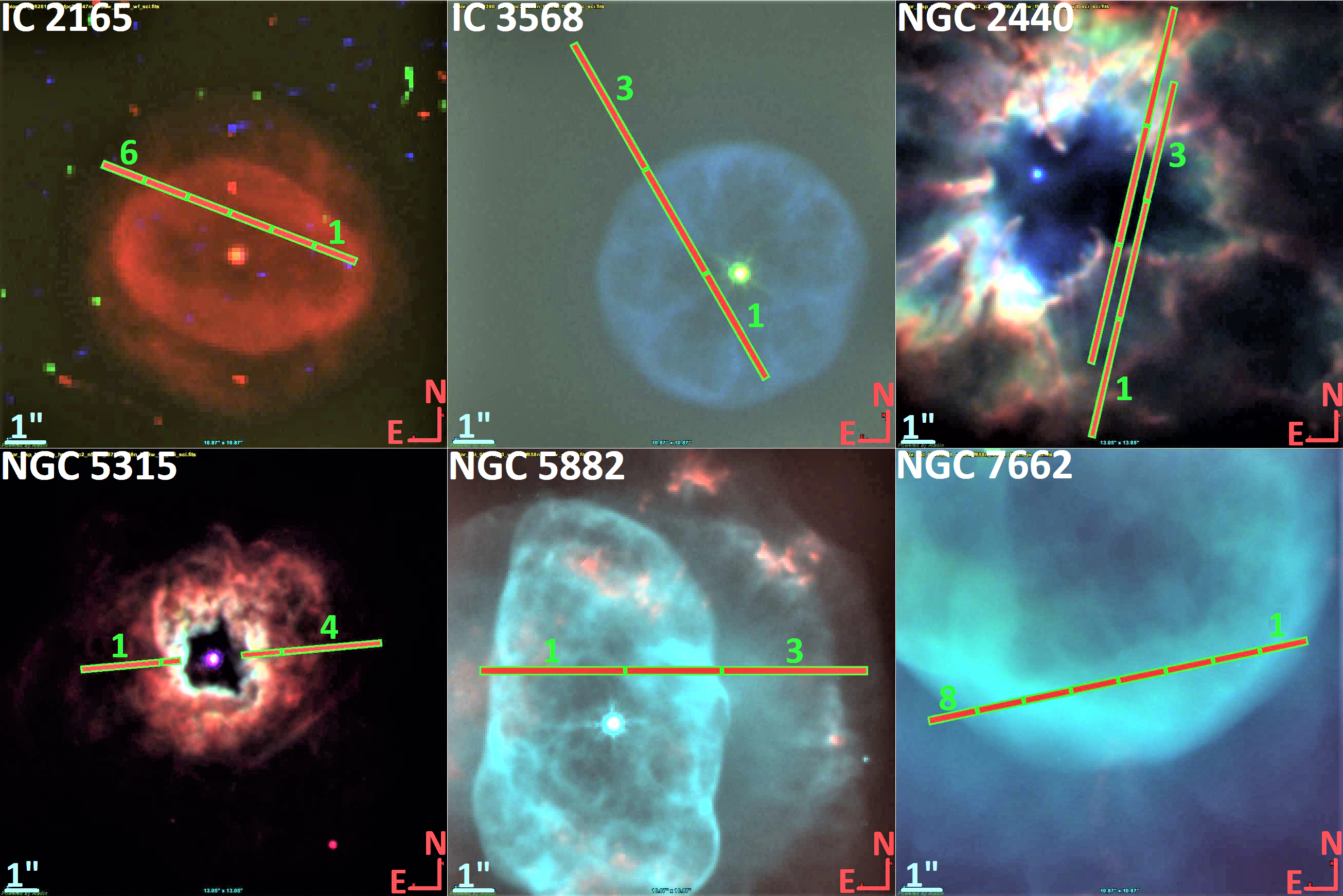}
\label{fig:slits}
\end{figure*}

A complete and accurate model of the evolution of stars occupying the mass range between 1 and 8 M$_{\astrosun}$ is particularly important for understanding the contribution that these stars make to the buildup of carbon and nitrogen in the interstellar medium. One requirement for a good model is the proper understanding of the mass ejection process, where some of the details of this process should be revealed by studying the spatial variations in the matter that comprises a planetary nebula (PN). In Miller et. al. (2016, hereafter Paper~III) the bright and multi-shelled planetary nebula, NGC 3242, was investigated. It was shown to have a uniform distribution of all observed elements, supporting current mixing length theories describing the transport of matter within the convective envelope and subsequent ejection from the star [see e.g. \citet{mar13} and \citet{ren81}]. In addition, through detailed modeling with Cloudy, the central star's temperature and luminosity were found to be 89.7$^{+7.3}_{-4.7}$~kK and log(L/$L_{\astrosun}$)=3.36$^{+0.28}_{-0.22}$, respectively. However, a single planetary nebula shown to be homogeneous is far from definitive. Here we analyze six additional PNe (IC 2165, IC 3568, NGC 2440, NGC 5315, NGC 5882, and NGC 7662) using the same procedures as those used in Paper~III with the purpose of addressing chemical homogeneity in planetary nebulae. 
\\\\The planetary nebulae in this study were first presented in Dufour et al. (2015; hereafter Paper~I), where the main goal was to measure critical emission lines of carbon and nitrogen for accurate abundance calculations of these elements. Among the selection criteria for the planetary nebulae were that each have a high surface brightness and, where practical, good angular size. These criteria make them ideal candidates for studying the chemical distribution of their matter. Each of the six planetary nebula can be seen in Figure~\ref{fig:slits}. IC 2165 is a slightly elliptical planetary nebula with shells detected at 2$''$, 4$''$, and 20$''$ (not visible in Figure~\ref{fig:slits}) from the central star~\citep{cor03}. IC 3568 has only two detected shells at radii of 4$''$ and 9$''$ (not visible in Figure~\ref{fig:slits}) and is very circular in appearance [\citet{bal92} and \citet{cor03}]. The shape of NGC 2440 is more complicated, with three bipolar structures emanating at various angles from the central star~\citep{lop98}, while NGC 5315 is an irregularly shaped, compact PN of radius 3$''$~\citep{ack92}. An elliptical shell with radial dimensions of 5.5$''$ x 3$''$, a more circular shell of radius 7.5$''$, and a halo (not visible in Figure~\ref{fig:slits}) that extends to 90$''$ are observed for NGC 5882 [\citet{cor00} and \citet{wel87}]. NGC 7662 is another triple-shell PN with two elliptical shells of radii 9$''$ x 6.2$''$ (only visible part in Figure~\ref{fig:slits}) and 15.4$''$ x 13.7$''$ encompassed by a 67$''$ circular shell~\citep{gue04}. 
Given these morphologies, the central star of each PN has experienced mass ejections \citep{sta95}, which may have been inhomogeneous in their elemental composition, either from shell to shell or across one ejection. Each shell could have a unique composition if each ejection occured with a different core temperature, resulting in a change in fusion rates around the core and convection to the surface. If the convection process was naturally asymmetrical or influenced by a companion star (stripping material), an ejection may result in an uneven distribution of elements. Therefore, the main goal of this fourth and final paper is to look for spatial differences in the abundances of carbon, nitrogen, helium, oxygen, neon, sulfur, and argon relative to hydrogen across each planetary nebula, using the spectroscopic data presented in Paper~I. 
\\\\ Turning to the problem of determining the central star properties, many authors have inferred the stellar temperature and luminosity using techniques such as photoionization modeling of the nebula (see e.g. Henry et al. 2015, hereafter Paper~II) and the Zanstra method [see e.g. Shaw and Kaler (1985, 1989);~\citet{zha93}]. Consequently, there is often a large range in the derived stellar parameters for any one object, resulting in uncertain initial/final masses for each star. This leads to our second goal; to constrain the stellar temperature and luminosity of each planetary nebula by modeling each with the photoionization code Cloudy, using the observations as constraints. Unlike previous photoionization modeling work, we use a physically motivated spectral energy distribution for the central star and calculate the modeled emission lines as seen through the slit instead of over the entire nebula.
\\\\The observations and analysis are presented in \S 2 followed by results in \S 3. Section 4 contains our discussion of the observational and modeling results while \S 5 contains the summary and conclusions.  

\section{Data Acquisition and Analysis}

Co-spatial HST/STIS spectra, with 0.05$''$ resolution and spanning the wavelength range of 1150-10,270~\AA, were obtained during the GO12600 spectra cycle 19 program and presented in Paper~I. The same methodology for spectral extraction and line analysis that was used in Paper~III was followed here. Specifically, a Python script was used to extract spectra for numerous regions in each PN, allowing for a comparison of abundances among regions. Each region's location was chosen to correspond with different structures within each nebula. The size of each region was chosen by looking at the signal-to-noise of the weakest lines and reducing the size until those lines were still able to be reliably measured. This maximized the total number of emission lines that could be used for analysis. All regions can be seen in Figure~\ref{fig:slits}, where the numbers identify each of the smaller regions, and the full region encompasses all of the numbered regions. The two slits shown for NGC 2440 were the result of a pointing error where the central star, which was going to be used as the guide star, was originally not within the field of view. Each slit shows the final UV (right) and optical (left) positions, so that the observations as a whole are not co-spatial for this one object. Emission line fluxes were measured with IRAF\footnotemark~using the task \textit{splot}. The same treatment in Paper~III for lines originating from the higher resolution gratings of STIS, namely summing the observed flux instead of fitting Gaussian profiles, was also used. This last operation corrected for the flat-top features of the emission lines exhibited in the higher resolution gratings. Any over-estimation of line strengths from the smaller regions was prevented by requiring the sum of the smaller regions' lines to equal the full region's value. The uncertainty estimates for each emission line were calculated by taking twice the FWHM of each emission line and multiplying it by the average of the continuum rms noise measured on either side of the line.
\footnotetext[1]{IRAF is distributed by the National Optical Astronomy Observatory, which is operated by the Association of Universities for Research in Astronomy (AURA) under cooperative agreement with the National Science Foundation.} 
\\\\The observed and dereddened line intensities for each region are shown in Table~\ref{emisstab1}. The first column specifies the wavelength measured in \AA, while the second column lists the ion responsible for the emission line. Each wavelength's value for the reddening function, f($\lambda$), can be found in the third column. The following pairs of columns contain the observed and dereddened line strengths under F($\lambda$) and I($\lambda$), respectively. All line strengths are normalized to F(H$\beta$) or I(H$\beta$) = 100. The dereddened values and accompanying errors were calculated using the program ELSA~\citep{joh06}. ELSA corrects the emission lines for interstellar extinction and contamination from He\textsuperscript{++} recombination lines while propagating the observed uncertainties. Following the last emission line of each region is the logarithmic reddening parameter, \textit{c}, the theoretical ratio of F(H$\alpha$/H$\beta$), and the observed, uncorrected flux of H$\beta$. Calculations for F(H$\alpha$/H$\beta$) occur in an iterative loop within ELSA since the value of \textit{c}, which is used to determine the ratio, depends on the nebular temperature and density. The lines that determine these nebular properties, typically [O III] lines for the temperature and C III] lines for the density, also depend on the value of \textit{c}, which necessitates an iterative loop in order to obtain a stable solution.
\begin{table*}
\centering
\caption{Fluxes and Intensities: The complete table is available online.}
\label{emisstab1}
\begin{tabular}{clccccccccccc}
\hline
Wave & & & \multicolumn{2}{c}{IC2165 Full} & \multicolumn{2}{c}{IC2165 1} & \multicolumn{2}{c}{IC2165 2} & \multicolumn{2}{c}{IC2165 3} & \multicolumn{2}{c}{IC2165 4} \\
\cline{4-13}
(\AA) & ID & f($\lambda$) & F($\lambda$) & I($\lambda$) & F($\lambda$) & I($\lambda$) & F($\lambda$) & I($\lambda$) & F($\lambda$) & I($\lambda$) & F($\lambda$) & I($\lambda$) \\
\hline
1485 & N IV] 	& 1.231 & 13.7 & 47.9$\pm$5.20 & 10.4 & 34.5$\pm$5.10 & 19.4 & 76.7$\pm$11.59 & 17.2 & 55.2$\pm$7.43 & 18.6 & 53.4$\pm$6.95 \\
1907 & C III] 	& 1.226 & 112 & 387$\pm$41 & 123 & 404$\pm$53 & 108 & 426$\pm$62 & 101 & 323$\pm$42 & 100 & 286$\pm$36 \\
1909 & C III] 	& 1.229 & 86.4 & 301$\pm$32 & 93.0 & 306$\pm$40 & 93.8 & 370$\pm$55 & 75.0 & 240$\pm$32 & 76.2 & 218$\pm$28 \\
3869 & [Ne III] 	& 0.224 & 59.4 & 76.8$\pm$5.40 & 65.8 & 84.1$\pm$6.29 & 48.3 & 64.0$\pm$7.00 & 47.7 & 60.5$\pm$7.18 & 44.6 & 55.3$\pm$6.24 \\
4363 & [O III] 	& 0.118 & 16.0 & 18.0$\pm$1.94 & 20.0 & 22.4$\pm$3.96 & 15.0 & 17.1$\pm$2.34 & 14.1 & 15.7$\pm$3.16 & 13.8 & 15.2$\pm$3.35 \\
4686 & He II 	& 0.036 & 60.0 & 62.2$\pm$2.97 & 58.4 & 60.5$\pm$3.59 & 80.5 & 83.8$\pm$4.75 & 68.3 & 70.7$\pm$4.14 & 78.7 & 81.2$\pm$3.70 \\
4861 & H$\beta$ 	& 0.000 & 100 & 100$\pm$0 & 100 & 100$\pm$0 & 100 & 100$\pm$0 & 100 & 100$\pm$0 & 100 & 100$\pm$0 \\
4959 & [O III] 	& -0.030 & 396 & 384$\pm$10 & 448 & 435$\pm$12 & 322 & 312$\pm$11 & 346 & 336$\pm$10 & 301 & 293$\pm$8 \\
5007 & [O III] 	& -0.042 & 1229 & 1178$\pm$26 & 1415 & 1359$\pm$36 & 980 & 936$\pm$28 & 1035 & 995$\pm$27 & 938 & 905$\pm$24 \\
5876 & He I 		& -0.231 & 10.6 & 8.41$\pm$1.18 & 11.2 & 8.92$\pm$2.53 & 9.48 & 7.32$\pm$2.99 & 8.29 & 6.66$\pm$2.13 & 6.37 & 5.23$\pm$1.91 \\
7136 & [Ar III] 	& -0.453 & 13.0 & 8.20$\pm$1.15 & 15.7 & 10.1$\pm$1.49 & 10.3 & 6.24$\pm$1.87 & 10.4 & 6.79$\pm$1.01 & 8.60 & 5.83$\pm$1.31 \\
9532 & [S III] 	& -0.632 & 50.5 & 26.6$\pm$1.57 & 45.9 & 24.9$\pm$2.92 & 45.8 & 22.6$\pm$2.22 & 47.1 & 25.9$\pm$3.01 & 36.6 & 21.3$\pm$2.92 \\
\hline
c\tablenotemark{a} & & & 0.44 & & 0.42 & & 0.48 & & 0.41 & & 0.37 \\
H$\alpha$/H$\beta$\tablenotemark{b} & & & 2.80 & & 2.79 & & 2.78 & & 2.80 & & 2.79 \\
log F$_{H\beta}$\tablenotemark{c} & & & -12.43 & & -13.11 & & -13.20 & & -13.21 & & -13.15 \\
\hline
\end{tabular}
\tablenotetext{a}{$\mathrm{L}$ogarithmic extinction at H$\beta$}
\tablenotetext{b}{Expected intrinsic H$\alpha$/H$\beta$ ratio at nebular temperature and density}
\tablenotetext{c}{ergs\ cm$^{-2}$ s$^{-1}$ in our extracted spectra}
\tablenotetext{d}{Blended 1907 and 1909}
\end{table*}

\section{Results}

\subsection{Element Distribution Within Each Planetary Nebula}
\label{sec:elem} 

\begin{figure*}
\caption{Helium, carbon, and nitrogen abundances (black points with solid error bar lines) for each region listed in Table~\ref{tab:abun1} scaled with respect to the full region of each PN. The abundances at nearly all positions in each object are consistent within errors, with outliers discussed in the text. Plotted just to the right of the measured abundances of each region are the model abundances also scaled with respect to the full region value (red points with dotted error bar lines).\newline}
\includegraphics[scale=0.405]{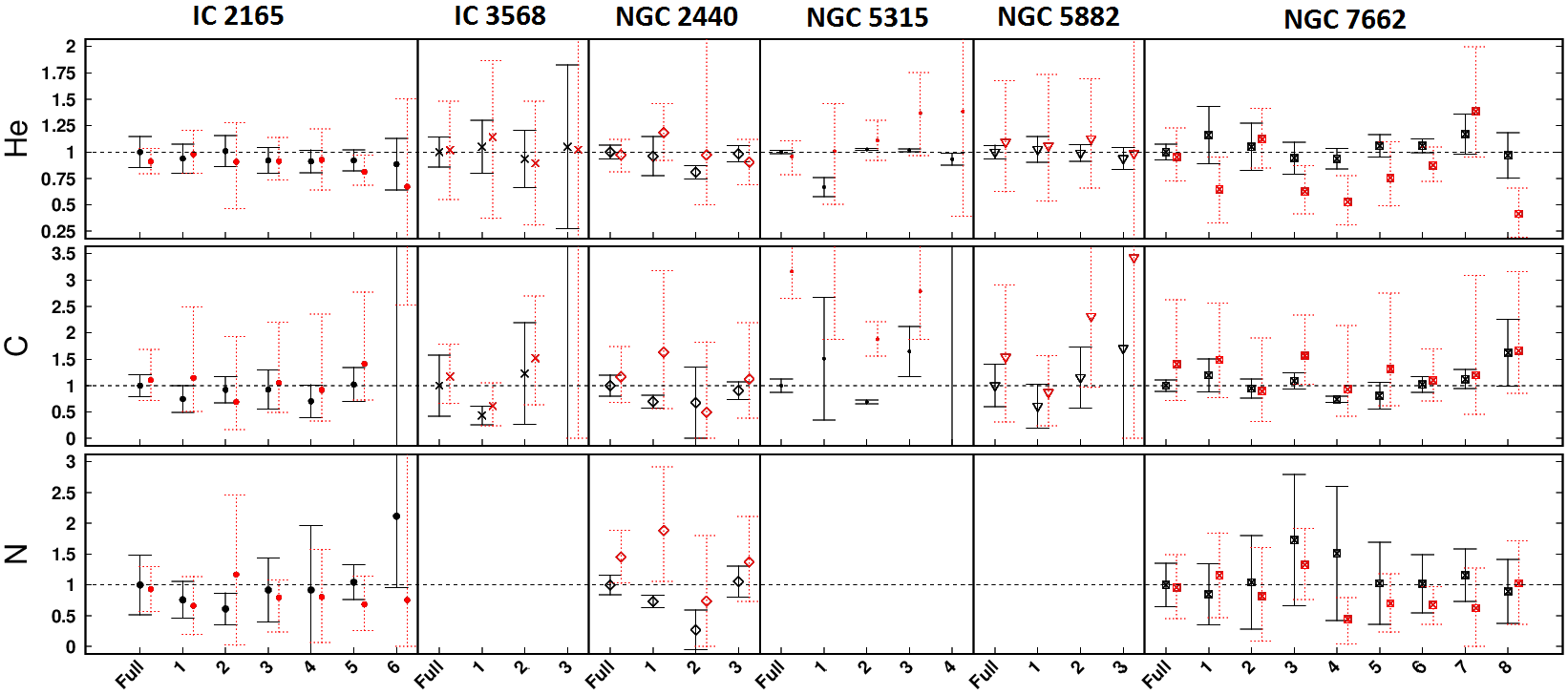}
\label{fig:abun1}
\end{figure*}

\begin{figure*}
\caption{Same as Figure~\ref{fig:abun1} but for oxygen, neon and argon.\newline}
\includegraphics[scale=0.405]{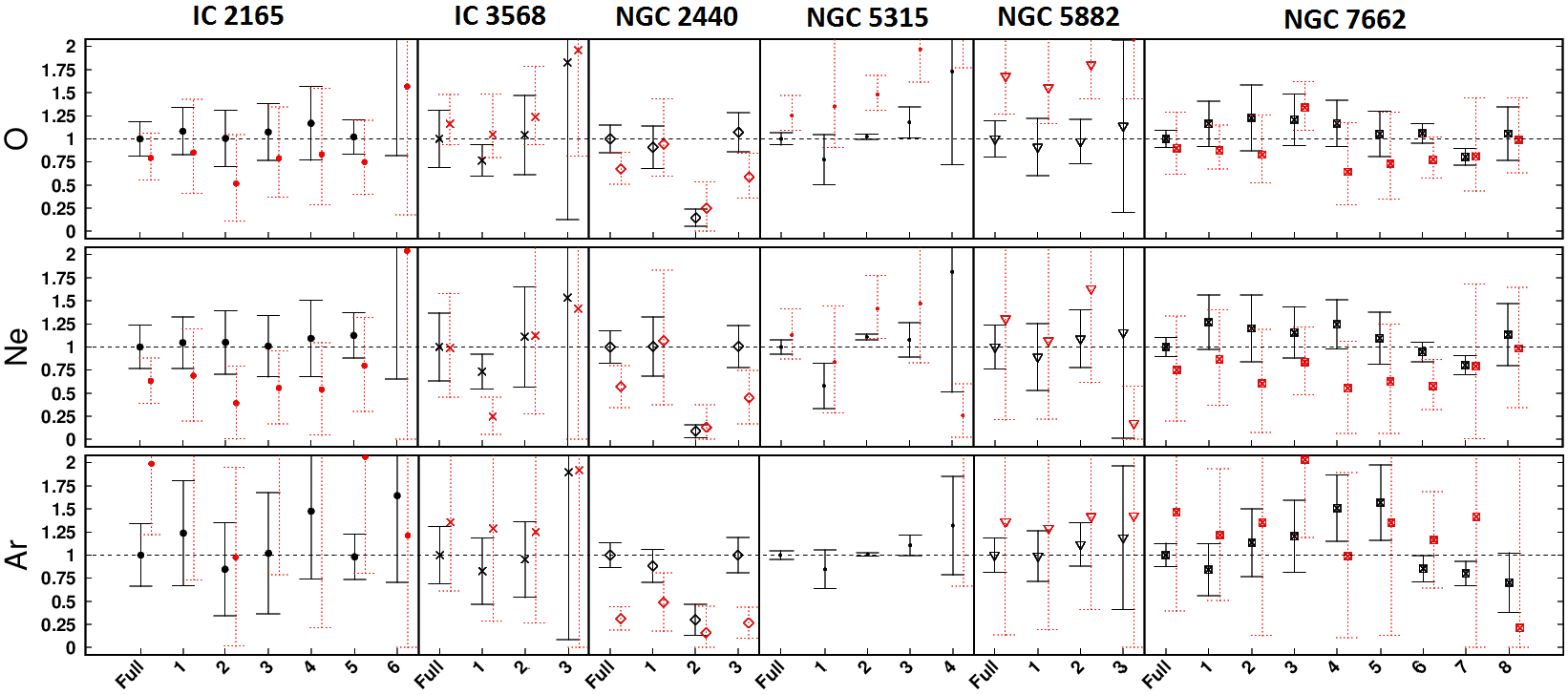}
\label{fig:abun2}
\end{figure*}

\begin{figure*}
\caption{Top panel: Same as Figure~\ref{fig:abun1} but for sulfur. Middle and bottom panels show the [O III] electron temperature and C III] electron density listed in Table~\ref{ionabun1} for each position in a PN.\newline}
\includegraphics[scale=0.405]{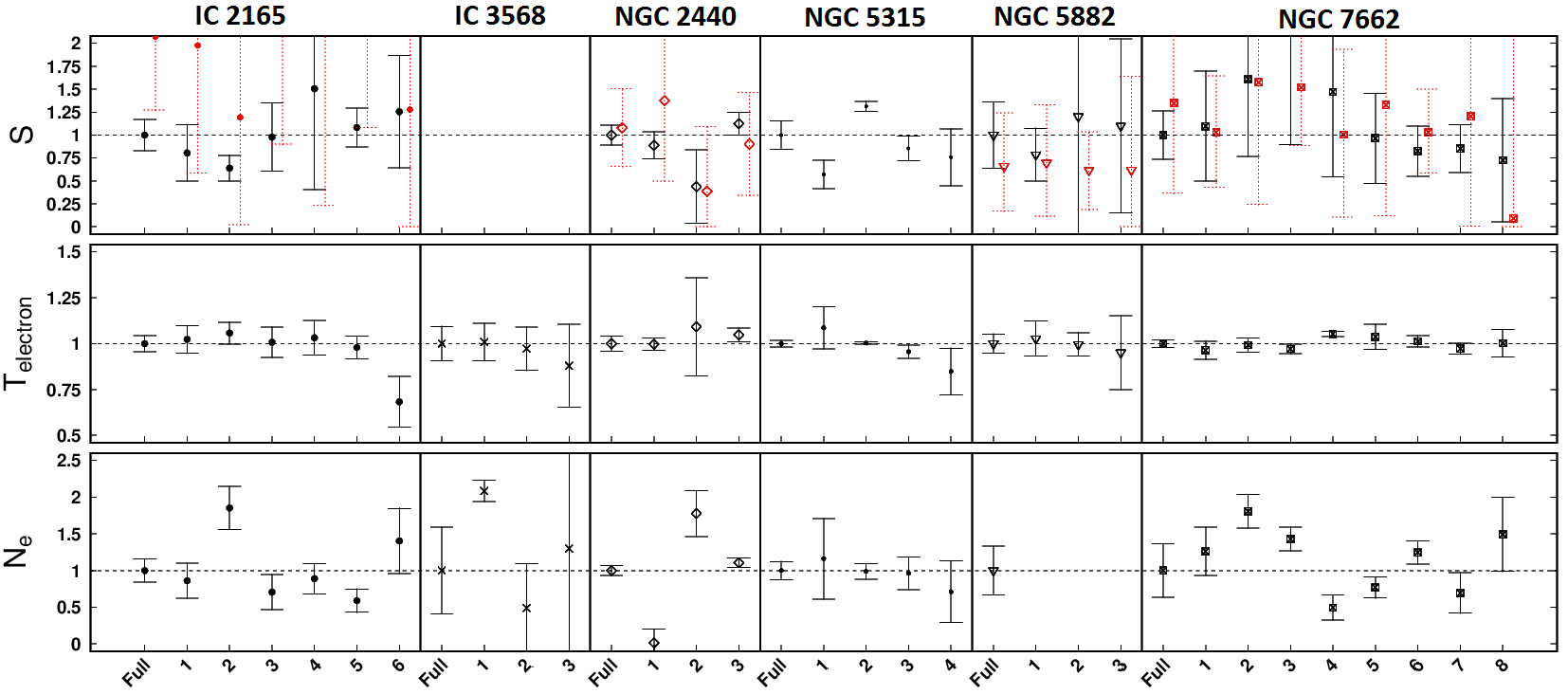}
\label{fig:abun3}
\end{figure*}

The nebular properties were also calculated using ELSA. To summarize the process, ELSA used the corrected line strengths and a 5-level atom scheme to calculate the abundances while propagating uncertainties of each ionic state. Then, for elements with unseen ionization states, ionization correction factors (ICFs) were calculated by ELSA and then applied to the sum of the observed ionic states to determine elemental abundances. These ICFs are calculated using the prescriptions outlined in Paper~I and \citet{kwi01}. For carbon and nitrogen, Paper~I showed that the direct sum of the observed ionic states was more accurate than an ICF derived total abundance so our total abundances for carbon and nitrogen are the result of summing the observed ionic states.
\\\\Table~\ref{ionabun1} lists the ionic abundances for every observed element. The ionic species and wavelength in \AA ngstroms of the specific emission line used to derive the ionic abundance are provided in the first column. Each column to the right of column 1 contains the ionic abundances for each region identified in the column head. Additionally, an ICF value is listed below the observed ions for each element (except for carbon and nitrogen, since those ICFs were not used to determine the total abundance). Each region follows in the remaining columns with its associated ionic abundances and ICFs. Finally, at the end of each object column are the inferred values for the temperature and density of the nebula.
\begin{table*}
\centering
\caption{Ionic Abundances\tablenotemark{a}, Temperatures and Densities: The complete table is available online.}
\label{ionabun1}
\begin{tabular}{cccccccc}
\hline
Ion & IC2165 Full & IC2165 1 & IC2165 2 & IC2165 3 & IC2165 4 & IC2165 5 & IC2165 6\\ 
\hline
He$^{+}$(5876) & 5.67$\pm$1.62(-2) & 5.12$\pm$1.50(-2) & 3.87$\pm$1.60(-2) & 3.95$\pm$1.30(-2) & 2.97$\pm$1.12(-2) & 6.86$\pm$1.07(-2) & 1.00$\pm$0.28(-1)\\
He$^{+2}$(4686) & 5.65$\pm$0.27(-2) & 5.47$\pm$0.33(-2) & 7.52$\pm$0.43(-2) & 6.42$\pm$0.39(-2) & 7.33$\pm$0.36(-2) & 3.58$\pm$0.28(-2) & \nodata\\
icf(He) & 1.00\tablenotemark{b} & 1.00\tablenotemark{b} & 1.00\tablenotemark{b} & 1.00\tablenotemark{b} & 1.00\tablenotemark{b} & 1.00\tablenotemark{b} & 1.00\tablenotemark{b}\\
O$^{+}$(3727) & 7.44$\pm$4.06(-6) & 7.91$\pm$3.45(-6) & 9.00$\pm$2.89(-6) & 5.80$\pm$3.51(-6) & 2.11$\pm$1.72(-6)  & 1.24$\pm$0.33(-5) & 1.48$\pm$0.77(-4)\\
O$^{+}$(7325) & 2.31$\pm$2.24(-5) & 8.55$\pm$4.08(-6) & 3.66$\pm$2.16(-6) & 9.65$\pm$8.20(-6) & 5.91$\pm$7.13(-6) & 2.02$\pm$0.79(-5) & 6.68$\pm$5.23(-5)\\
O$^{+}$(adopt) & 1.14$\pm$0.41(-5) & 8.06$\pm$2.76(-6) & 7.94$\pm$2.54(-6) & 6.90$\pm$3.60(-6) & 3.74$\pm$4.18(-6) & 1.44$\pm$0.33(-5) & 1.29$\pm$0.69(-4)\\ 
O$^{+2}$(5007) & 1.60$\pm$0.21(-4) & 1.71$\pm$0.37(-4) & 1.08$\pm$0.18(-4) & 1.31$\pm$0.32(-4) & 1.11$\pm$0.30(-4) & 2.13$\pm$0.41(-4) & 7.61$\pm$5.41(-4)\\
O$^{+2}$(4959) & 1.55$\pm$0.20(-4) & 1.64$\pm$0.35(-4) & 1.08$\pm$0.18(-4) & 1.32$\pm$0.32(-4) & 1.08$\pm$0.29(-4) & 2.15$\pm$0.42(-4) & 6.60$\pm$4.69(-4)\\
O$^{+2}$(4363) & 1.56$\pm$0.21(-4) & 1.71$\pm$0.37(-4) & 1.08$\pm$0.18(-4) & 1.31$\pm$0.32(-4) & 1.11$\pm$0.30(-4) & 2.13$\pm$0.41(-4) & 7.61$\pm$5.41(-4)\\
O$^{+2}$(adopt) & 1.59$\pm$0.20(-4) & 1.70$\pm$0.37(-4) & 1.08$\pm$0.18(-4) & 1.32$\pm$0.32(-4) & 1.10$\pm$0.30(-4) & 2.13$\pm$0.41(-4) & 7.39$\pm$5.25(-4)\\
icf(O) & 2.00$\pm$0.29 & 2.07$\pm$0.31 & 2.94$\pm$0.80 & 2.63$\pm$0.54 & 3.47$\pm$0.92 & 1.52$\pm$0.09 & 1.00\tablenotemark{b}\\
Ar$^{+2}$(7135) & 3.93$\pm$0.67(-7) & 4.59$\pm$1.01(-7) & 2.65$\pm$0.86(-7) & 3.20$\pm$0.75(-7) & 2.61$\pm$0.79(-7) & 5.04$\pm$1.00(-7) & 1.30$\pm$0.72(-6)\\
Ar$^{+2}$(7751) & 6.94$\pm$3.54(-7) & 8.61$\pm$5.81(-7) & 3.40$\pm$4.71(-7) & 5.37$\pm$5.92(-7) & 6.76$\pm$3.36(-7) & 8.87$\pm$3.35(-7) & 1.70$\pm$1.58(-6)\\
Ar$^{+2}$(adopt) & 4.88$\pm$1.52(-7) & 5.91$\pm$2.63(-7) & 2.84$\pm$1.53(-7) & 3.86$\pm$2.39(-7) & 4.27$\pm$1.90(-7) & 6.25$\pm$1.60(-7) & 1.41$\pm$0.84(-6)\\ 
icf(Ar) & 2.07$\pm$0.28 & 2.11$\pm$0.32 & 3.02$\pm$0.81 & 2.68$\pm$0.54 & 3.50$\pm$0.92 & 1.59$\pm$0.09 & 1.18$\pm$0.05\\
C$^{+}$(2325) & 1.00$\pm$0.16(-6) & 7.56$\pm$1.81(-7) & 8.29$\pm$2.71(-7) & 5.24$\pm$2.48(-7) & 4.09$\pm$1.56(-7) & 1.17$\pm$0.26(-6) & 1.08$\pm$0.40(-5)\\
C$^{+2}$(1909) & 2.14$\pm$0.58(-5) & 1.91$\pm$0.84(-4) & 1.77$\pm$0.61(-4) & 1.66$\pm$0.83(-4) & 1.29$\pm$0.71(-4) & 2.63$\pm$1.07(-4) & 5.69$\pm$9.45(-5)\\
C$^{+2}$(1907) & 2.14$\pm$0.58(-5) & 1.90$\pm$0.84(-4) & 1.76$\pm$0.61(-4) & 1.66$\pm$0.83(-4) & 1.29$\pm$0.71(-4) & 2.63$\pm$1.06(-4) & 5.69$\pm$9.44(-5)\\
C$^{+2}$(adopt) & 2.14$\pm$0.58(-5) & 1.90$\pm$0.84(-4) & 1.76$\pm$0.61(-4) & 1.66$\pm$0.83(-4) & 1.29$\pm$0.71(-4) & 2.63$\pm$1.07(-4) & 5.69$\pm$9.44(-4)\\
C$^{+3}$(1549) & 2.00$\pm$0.65(-5) & 1.19$\pm$0.63(-4) & 2.05$\pm$0.83(-4) & 2.17$\pm$1.30(-4) & 1.62$\pm$1.07(-4) & 1.60$\pm$0.78(-4) & 3.21$\pm$6.54(-3)\\
N$^{+}$(6584) & 3.80$\pm$0.28(-5) & 1.91$\pm$0.19(-6) & 1.66$\pm$0.15(-6) & 1.95$\pm$0.22(-6) & 9.01$\pm$1.28(-7) & 4.84$\pm$0.40(-6) & 3.60$\pm$1.25(-5)\\
N$^{+}$(6548) & 3.46$\pm$0.33(-5) & 1.65$\pm$0.28(-6) & 1.26$\pm$0.28(-6) & 1.85$\pm$0.32(-6) & 1.12$\pm$0.28(-6) & 4.75$\pm$0.49(-6) & 3.14$\pm$1.10(-5)\\
N$^{+}$(adopt) & 3.72$\pm$0.28(-5) & 1.85$\pm$0.19(-6) & 1.58$\pm$0.14(-6) & 1.93$\pm$0.22(-6) & 9.67$\pm$1.48(-7) & 4.82$\pm$0.40(-6) & 3.49$\pm$1.21(-5)\\
N$^{+3}$(1485) & 6.80$\pm$2.28(-5) & 4.07$\pm$2.26(-5) & 7.08$\pm$3.00(-5) & 7.38$\pm$4.61(-5) & 5.92$\pm$4.08(-5) & 4.99$\pm$2.56(-5) & 2.09$\pm$4.54(-4)\\
N$^{+4}$(1240) & 4.62$\pm$1.84(-5) & 1.04$\pm$0.75(-5) & 4.81$\pm$2.40(-5) & 7.30$\pm$5.36(-5) & 4.45$\pm$3.60(-5) & 1.18$\pm$0.82(-5) & \nodata\\ 
Ne$^{+2}$(3869) & 2.57$\pm$0.41(-5) & 2.59$\pm$0.65(-6) & 1.77$\pm$0.37(-5) & 1.97$\pm$0.59(-5) & 1.66$\pm$0.54(-5) & 3.75$\pm$0.85(-5) & 1.64$\pm$1.40(-4)\\
Ne$^{+3}$(1602) & 6.59$\pm$2.27(-5) & \nodata & \nodata & \nodata & \nodata & \nodata & \nodata\\
Ne$^{+4}$(1575) & 1.04$\pm$0.37(-5) & \nodata & \nodata & \nodata & \nodata & \nodata & \nodata\\
icf(Ne) & 2.09$\pm$0.35 & 2.16$\pm$0.34 & 3.19$\pm$0.88  & 2.74$\pm$0.57 & 3.53$\pm$0.95 & 1.61$\pm$0.10 & 1.20$\pm$0.07\\
S$^{+}$(6716) & 1.04$\pm$0.62(-7) & \nodata & \nodata & \nodata & \nodata & \nodata & \nodata\\
S$^{+}$(6731) & 1.01$\pm$0.63(-7) & \nodata & \nodata & \nodata & \nodata & \nodata & \nodata\\
S$^{+}$(adopt) & 1.02$\pm$0.62(-7) & \nodata & \nodata & \nodata & \nodata & \nodata & \nodata\\
S$^{+2}$(9069) & \nodata & 7.33$\pm$2.33(-7) & \nodata & \nodata & \nodata & \nodata & \nodata\\
S$^{+2}$(9532) & 8.13$\pm$0.86(-7) & \nodata & 6.25$\pm$0.94(-7) & 7.8$\pm$1.58(-7) & 6.14$\pm$1.41(-7) & 1.18$\pm$0.18(-6) & 1.85$\pm$0.85(-6)\\
icf(S) & 1.97$\pm$0.34 & 1.98$\pm$0.35 & 1.84$\pm$0.26 & 2.26$\pm$0.66 & 4.42$\pm$2.94 & 1.66$\pm$0.14 & 1.22$\pm$0.05\\
\hline
[O III] T$_{e}$(K) & 13600$\pm$600 & 13900$\pm$1000 & 14300$\pm$800 & 13700$\pm$1100 & 14000$\pm$1300 & 13300$\pm$900 & 9300$\pm$1900\\
$ $[S II] N$_{e}$($cm^{-3}$) & 1500$\pm$3800 & \nodata & \nodata & \nodata & \nodata & \nodata & \nodata\\
C III] N$_{e}$($cm^{-3}$) & 7700$\pm$1200 & 6600$\pm$1800 & 14200$\pm$2300 & 5400$\pm$1800 & 6800$\pm$1600 & 4500$\pm$1200 & 10800$\pm$3400\\
\hline
\end{tabular}
\tablenotetext{a}{\footnotesize Abundances relative to H$^{+}$; n.nn$\pm$n.nn(-k) == (n.nn$\pm$n.nn) x 10$^{-k}$}
\tablenotetext{b}{\footnotesize Assumed value}
\end{table*}
The final abundances for each of the seven elements measured in each region are presented in Table~\ref{tab:abun1}. The abundances relative to hydrogen or oxygen are given in columns 2-7, while the solar values from~\citet{asp09} are provided in column 8. Figures~\ref{fig:abun1}-\ref{fig:abun3} compare the total abundances of each region with the values scaled with respect to the abundance of the full region in each planetary nebula. The bottom two panels of figure~\ref{fig:abun3} also compare the [O III] temperature and C III] density of each region with each value scaled in the same manner as the abundances. Out of all the regions, only region 2 of NGC 2440 and region 1 of NGC 5315 show abundances that are significantly lower than the respective full regions. For region 2 of NGC 2440, this can likely be attributed to the fact that the ICF for oxygen is only known to be greater than 1. All of the abundances except for helium, carbon, and nitrogen, depend on this ICF. The abundances shown for oxygen, neon, argon, and sulfur assume the ICF of oxygen equals 1, and therefore, these abundances are likely underestimated. For region 1 of NGC 5315, a higher electron temperature is measured compared with the temperature of its full region, causing the inferred abundances to be lower. However, the temperature is consistent (given the errors) with the full region, and substituting the full region's temperature value in for the abundance calculations of region 1 yields abundances consistent with the full region's values.
\begin{table*}
\centering
\caption{Total Elemental Abundances: The complete table is available online.}
\label{tab:abun1}
\begin{tabular}{lcccccccc}
\hline
Parameter & IC2165 Full & IC2165 1 & IC2165 2 & IC2165 3 & IC2165 4 & IC2165 5 & IC2165 6 & Solar\tablenotemark{a}\\ 
\hline
He/H (10$^{-2}$) 	& 11.30$^{+1.65}_{-1.65}$ & 10.60$^{+1.56}_{-1.56}$ & 11.40$^{+1.67}_{-1.67}$ & 10.40$^{+1.38}_{-1.38}$ & 10.30$^{+1.21}_{-1.21}$ & 10.40$^{+1.13}_{-1.13}$ & 10.00$^{+2.78}_{-2.78}$ & 8.51 \\
C/H (10$^{-4}$) 	& 4.14$^{+0.87}_{-0.87}$ & 3.09$^{+1.05}_{-1.05}$ & 3.81$^{+1.03}_{-1.03}$ & 3.83$^{+1.54}_{-1.54}$ & 2.91$^{+1.28}_{-1.28}$ & 4.23$^{+1.32}_{-1.32}$ & 89.00$^{+114.84}_{-114.84}$ & 2.69 \\
C/O 			& 1.221$^{+0.343}_{-0.343}$ & 0.842$^{+0.348}_{-0.348}$ & 1.117$^{+0.453}_{-0.453}$ & 1.052$^{+0.519}_{-0.519}$ & 0.735$^{+0.409}_{-0.409}$ & 1.223$^{+0.443}_{-0.443}$ & 10.253$^{+14.954}_{-14.954}$ & 0.550 \\
N/H (10$^{-4}$) 	& 1.18$^{+0.29}_{-0.29}$ & 0.53$^{+0.24}_{-0.24}$ & 1.21$^{+0.38}_{-0.38}$ & 1.49$^{+0.71}_{-0.71}$ & 1.05$^{+0.54}_{-0.54}$ & 0.67$^{+0.27}_{-0.27}$ & 2.44$^{+4.54}_{-4.54}$ & 0.68 \\
N/O 			& 0.348$^{+0.108}_{-0.108}$ & 0.144$^{+0.073}_{-0.073}$ & 0.353$^{+0.155}_{-0.155}$ & 0.409$^{+0.227}_{-0.227}$ & 0.264$^{+0.164}_{-0.164}$ & 0.192$^{+0.085}_{-0.085}$ & 0.281$^{+0.557}_{-0.557}$ & 0.138 \\
O/H (10$^{-4}$) 	& 3.39$^{+0.64}_{-0.64}$ & 3.67$^{+0.87}_{-0.87}$ & 3.41$^{+1.03}_{-1.03}$ & 3.64$^{+1.04}_{-1.04}$ & 3.96$^{+1.35}_{-1.35}$ & 3.46$^{+0.63}_{-0.63}$ & 8.68$^{+5.90}_{-5.90}$ & 4.90 \\
Ne/H (10$^{-5}$) 	& 5.36$^{+1.26}_{-1.26}$ & 5.61$^{+1.49}_{-1.49}$ & 5.63$^{+1.84}_{-1.84}$ & 5.41$^{+1.78}_{-1.78}$ & 5.86$^{+2.22}_{-2.22}$ & 6.03$^{+1.31}_{-1.31}$ & 19.50$^{+16.00}_{-16.00}$ & 8.51 \\
Ne/O 			& 0.158$^{+0.048}_{-0.048}$ & 0.153$^{+0.054}_{-0.054}$ & 0.165$^{+0.073}_{-0.073}$ & 0.149$^{+0.065}_{-0.065}$ & 0.148$^{+0.075}_{-0.075}$ & 0.174$^{+0.049}_{-0.049}$ & 0.225$^{+0.239}_{-0.239}$ & 0.174 \\
S/H (10$^{-6}$) 	& 1.80$^{+0.31}_{-0.31}$ & 1.45$^{+0.56}_{-0.56}$ & 1.15$^{+0.25}_{-0.25}$ & 1.76$^{+0.67}_{-0.67}$ & 2.71$^{+1.98}_{-1.98}$ & 1.95$^{+0.38}_{-0.38}$ & 2.26$^{+1.10}_{-1.10}$ & 13.2 \\
S/O 			& 0.005$^{+0.001}_{-0.001}$ & 0.004$^{+0.002}_{-0.002}$ & 0.003$^{+0.001}_{-0.001}$ & 0.005$^{+0.002}_{-0.002}$ & 0.007$^{+0.006}_{-0.006}$ & 0.006$^{+0.002}_{-0.002}$ & 0.003$^{+0.002}_{-0.002}$ & 0.027 \\
Ar/H (10$^{-7}$) 	& 10.10$^{+3.42}_{-3.42}$ & 12.50$^{+5.73}_{-5.73}$ & 8.55$^{+5.08}_{-5.08}$ & 10.30$^{+6.62}_{-6.62}$ & 14.90$^{+7.39}_{-7.39}$ & 9.91$^{+2.48}_{-2.48}$ & 16.60$^{+9.47}_{-9.47}$ & 25.12 \\
Ar/O 			& 0.003$^{+0.001}_{-0.001}$ & 0.003$^{+0.002}_{-0.002}$ & 0.003$^{+0.002}_{-0.002}$ & 0.003$^{+0.002}_{-0.002}$ & 0.004$^{+0.002}_{-0.002}$ & 0.003$^{+0.001}_{-0.001}$ & 0.002$^{+0.002}_{-0.002}$ & 0.005 \\
\hline
\end{tabular}
\tablenotetext{a}{\citet{asp09}}
\end{table*}
\\\\A literature search over the last twenty years shows that only NGC 2440 and NGC 5882 from our sample have been investigated previously for abundance variations.~\citet{per98} presented results on the abundances of helium, oxygen, nitrogen, neon, argon, and sulfur for NGC 2440. They made long-slit observations from 3600-9600 \AA~along the major axis using the Boller \& Chivens spectrograph and the 1.52m ESO telescope at La Silla, Chile. The observations were divided into seven regions spanning the slit length and width of 76.1$''$ and 2$''$, respectively. There were no observed spatial variations between the seven regions within their uncertainties for every element. NGC 5882 was observed by~\citet{gue99} over the wavelength range 3400-10,300 \AA~using the same instrument as the one employed by~\citet{per98}. They compared the outermost halo to the inner two shells as a whole. They found the abundances of helium, oxygen, and neon to be consistent within errors between the halo and shells. 
\\\\Table~\ref{tab:abuncomp} compares the full region abundances for each planetary nebula with other values in the literature. Each PN has two comparison values from either authors with quoted uncertainties or average values with standard deviations from multiple authors. The majority of the abundances match within 1$\sigma$ and nearly all are consistent within 2$\sigma$ with the other authors. The largest discrepancy is seen in the nitrogen abundance of NGC 2440 when compared to the value of \citet{kwi03}. This is most likely caused by the fact that the nitrogen abundance of \citet{kwi03} was calculated using the ICF method (ICF value of 9.71) whereas our value is the sum of the observed ionic species, which includes the most populated ionic states. 
\begin{table*}
\centering
\caption{Comparison: Total Elemental Abundances}
\label{tab:abuncomp}
\begin{tabular}{lcccccc}
\hline
 & \multicolumn{3}{c}{IC 2165 Full} & \multicolumn{3}{c}{IC 3568 Full}\\
\hline
Element & This Work & \citet{kwi03} & BH\tablenotemark{a} & This Work & \citet{hen04} & \citet{kwi98} \\
\hline
He/H (10$^{-2}$) 	& 11.30$^{+1.65}_{-1.65}$ & 9.00$^{+3.00}_{-3.00}$ & 10.83$^{+0.66}_{-0.66}$ & 9.53$^{+1.36}_{-1.36}$ & 12.00$^{+4.00}_{-4.00}$ & 11.00$^{+1.65}_{-1.65}$ \\
C/H (10$^{-4}$) 	& 4.14$^{+0.87}_{-0.87}$ & \nodata & 6.54$^{+1.76}_{-1.76}$ & 4.66$^{+2.71}_{-2.71}$ & \nodata & 0.97$^{+0.49}_{-0.49}$ \\
N/H (10$^{-4}$) 	& 1.18$^{+0.29}_{-0.29}$ & 1.33$^{+0.19}_{-0.19}$ & 1.20$^{+0.32}_{-0.32}$ & \nodata & \nodata & \nodata\\
O/H (10$^{-4}$) 	& 3.39$^{+0.64}_{-0.64}$ & 3.11$^{+0.44}_{-0.44}$ & 2.28$^{+0.31}_{-0.31}$ & 4.06$^{+1.26}_{-1.26}$ & 3.77$^{+0.53}_{-0.53}$ & 3.14$^{+0.47}_{-0.47}$ \\
Ne/H (10$^{-5}$) 	& 5.36$^{+1.26}_{-1.26}$ & 6.90$^{+0.93}_{-0.93}$ & 5.66$^{+2.23}_{-2.23}$ & 8.15$^{+1.26}_{-1.26}$ & 7.16$^{+1.13}_{-1.13}$ & 4.08$^{+0.61}_{-0.61}$ \\
S/H (10$^{-6}$) 	& 1.80$^{+0.31}_{-0.31}$ & 2.70$^{+0.31}_{-0.31}$ & 3.62$^{+1.29}_{-1.29}$ & \nodata & \nodata & \nodata\\
Ar/H (10$^{-7}$) 	& 10.10$^{+3.42}_{-3.42}$ & 16.00$^{+2.18}_{-2.18}$ & 13.23$^{+4.57}_{-4.57}$ & 12.70$^{+3.91}_{-3.91}$ & 10.56$^{+3.39}_{-3.39}$ & \nodata\\
\hline
 & \multicolumn{3}{c}{NGC 2440 Full} & \multicolumn{3}{c}{NGC 5315 Full}\\
\hline
Element & This Work & \citet{kwi03} & \citet{hyu98} & This Work & \citet{pei04} & TKSDT\tablenotemark{b}\\
\hline
He/H (10$^{-2}$) 	& 12.60$^{+0.84}_{-0.84}$ & 10.00$^{+3.00}_{-3.00}$ & 12.00$^{+1.20}_{-1.20}$ & 11.90$^{+0.17}_{-0.17}$ & 11.75$^{+0.11}_{-0.11}$ & 10.71$^{+1.20}_{-1.20}$\\
C/H (10$^{-4}$) 	& 2.18$^{+0.44}_{-0.44}$ & \nodata & 4.00$^{+2.00}_{-2.00}$ & 5.20$^{+0.65}_{-0.65}$ & 7.08$^{+1.47}_{-1.47}$ & 5.96$^{+5.40}_{-5.40}$\\
N/H (10$^{-4}$) 	& 4.62$^{+0.77}_{-0.77}$ & 10.67$^{+1.51}_{-1.51}$ & 9.50$^{+2.38}_{-2.38}$ & 7.86$^{+1.01}_{-1.01}$ & 6.92$^{+2.39}_{-2.39}$ & 3.61$^{+1.75}_{-1.75}$\\
O/H (10$^{-4}$) 	& 4.30$^{+0.65}_{-0.65}$ & 5.19$^{+0.73}_{-0.73}$ & 4.40$^{+0.70}_{-0.70}$ & 5.04$^{+0.33}_{-0.33}$ & 7.41$^{+0.51}_{-0.51}$ & 6.02$^{+1.55}_{-1.55}$\\
Ne/H (10$^{-5}$) 	& 7.03$^{+1.24}_{-1.24}$ & 10.70$^{+1.56}_{-1.56}$ & 9.00$^{+1.35}_{-1.35}$ & 12.90$^{+1.01}_{-1.01}$ & 20.41$^{+4.70}_{-4.70}$ & 16.66$^{+4.95}_{-4.95}$\\
S/H (10$^{-6}$) 	& 1.52$^{+0.16}_{-0.16}$ & 2.80$^{+0.52}_{-0.52}$ & 2.00$^{+0.60}_{-0.60}$ & 20.70$^{+3.23}_{-3.23}$ & 32.35$^{+8.94}_{-8.94}$ & 17.87$^{+3.31}_{-3.31}$\\
Ar/H (10$^{-7}$) 	& 48.00$^{+6.47}_{-6.47}$ & 40.60$^{+5.71}_{-5.71}$ & 21.00$^{+4.20}_{-4.20}$ & 37.80$^{+1.70}_{-1.70}$ & 56.21$^{+10.35}_{-10.35}$ & 49.58$^{+14.20}_{-14.21}$\\
\hline
& \multicolumn{3}{c}{NGC 5882 Full} & \multicolumn{3}{c}{NGC 7662 Full}\\
\hline
Element & This Work & \citet{kwi03} & \citet{gue99} & This Work & \citet{kwi03} & LHBA\tablenotemark{c} \\
\hline
He/H (10$^{-2}$) 	& 11.50$^{+0.73}_{-0.73}$ & 12.00$^{+4.00}_{-4.00}$ & 11.10$^{+0.70}_{-0.70}$ & 10.00$^{+0.76}_{-0.76}$ & 10.00$^{+3.00}_{-3.00}$ & 10.05$^{+1.15}_{-1.15}$\\
C/H (10$^{-4}$) 	& 2.72$^{+1.09}_{-1.09}$ & \nodata & \nodata & 4.08$^{+0.43}_{-0.43}$ & \nodata & 6.27$^{+2.48}_{-2.48}$\\
N/H (10$^{-4}$) 	& 2.34$^{+0.83}_{-0.83}$ & 1.80$^{+0.27}_{-0.27}$ & 1.40$^{+0.40}_{-0.40}$ & 0.77$^{+0.11}_{-0.11}$ & 0.74$^{+0.13}_{-0.13}$ & 1.00$^{+0.42}_{-0.42}$\\
O/H (10$^{-4}$) 	& 7.63$^{+1.49}_{-1.49}$ & 5.48$^{+0.77}_{-0.77}$ & 4.93$^{+0.45}_{-0.45}$ & 4.47$^{+0.42}_{-0.42}$ & 4.19$^{+0.59}_{-0.59}$ & 4.59$^{+1.70}_{-1.70}$\\
Ne/H (10$^{-5}$) 	& 17.80$^{+4.21}_{-4.21}$ & 14.80$^{+2.19}_{-2.19}$ & 8.40$^{+1.20}_{-1.20}$ & 8.99$^{+0.92}_{-0.92}$ & 7.10$^{+0.84}_{-0.84}$ & 7.58$^{+1.80}_{-1.80}$\\
S/H (10$^{-6}$) 	& 11.90$^{+4.29}_{-4.29}$ & 7.20$^{+1.10}_{-1.10}$ & 7.00$^{+0.70}_{-0.70}$ & 2.59$^{+0.68}_{-0.68}$ & 4.30$^{+0.42}_{-0.42}$ & 5.71$^{+1.52}_{-1.52}$\\
Ar/H (10$^{-7}$) 	& 28.50$^{+5.34}_{-5.34}$ & 27.90$^{+3.84}_{-3.84}$ & 31.50$^{+1.08}_{-1.08}$ & 17.10$^{+2.16}_{-2.16}$ & 24.60$^{+3.35}_{-3.35}$ & 14.80$^{+4.46}_{-4.46}$\\
\hline
\end{tabular}
\tablenotetext{a}{Values and uncertainties are averages and standard deviations, respectively, from \citet{boh13} and \citet{hyu94}.}
\tablenotetext{b}{Values and uncertainties are averages and standard deviations, respectively, from \citet{tsa03}, \citet{kin94}, \citet{sam92}, \citet{def91}, and \citet{tor81}.}
\tablenotetext{c}{Values and uncertainties are averages and standard deviations, respectively, from \citet{liu04}, \citet{hyu97}, \citet{bar86}, and \citet{all83}.}
\end{table*}
Looking at the temperatures of each region, only region 6 of IC 2165 is significantly lower than its full region temperature. This region probes the middle shell of IC 2165, as opposed to the other regions that probe the innermost shell. Therefore, this may indeed be cooler gas since it is farther from the central star and may be optically thick such that the effect of heating by radiation hardening doesn't occur. Lastly, each of the density variations appears to coincide with brighter areas of each PN for the higher densities, and dimmer areas for the lower densities. 
\\\\To summarize, first, we presented dereddened emission line intensities for various regions across each planetary nebula. Next, these emission lines were used to calculate nebular properties of temperature and density as well as ionic abundances. Last, total abundances for He, C, N, O, Ne, S and Ar were calculated from the ionic abundances for each region. \textit{The major result is that each of the six planetary nebulae appears to exhibit a chemically homogeneous distribution of these elements.}

\subsection{Central Star Properties}
\label{sec:star} 

\begin{figure*}
\caption{The three different density profiles used for the full regions of each planetary nebula are shown here. The solid red lines are the constant density profiles. The blue dash lines are the Gaussian density profile. The combination of the blue dash lines and green dot-dash lines are the Gaussian with a power-law density profiles.\newline}
\includegraphics[scale=0.235]{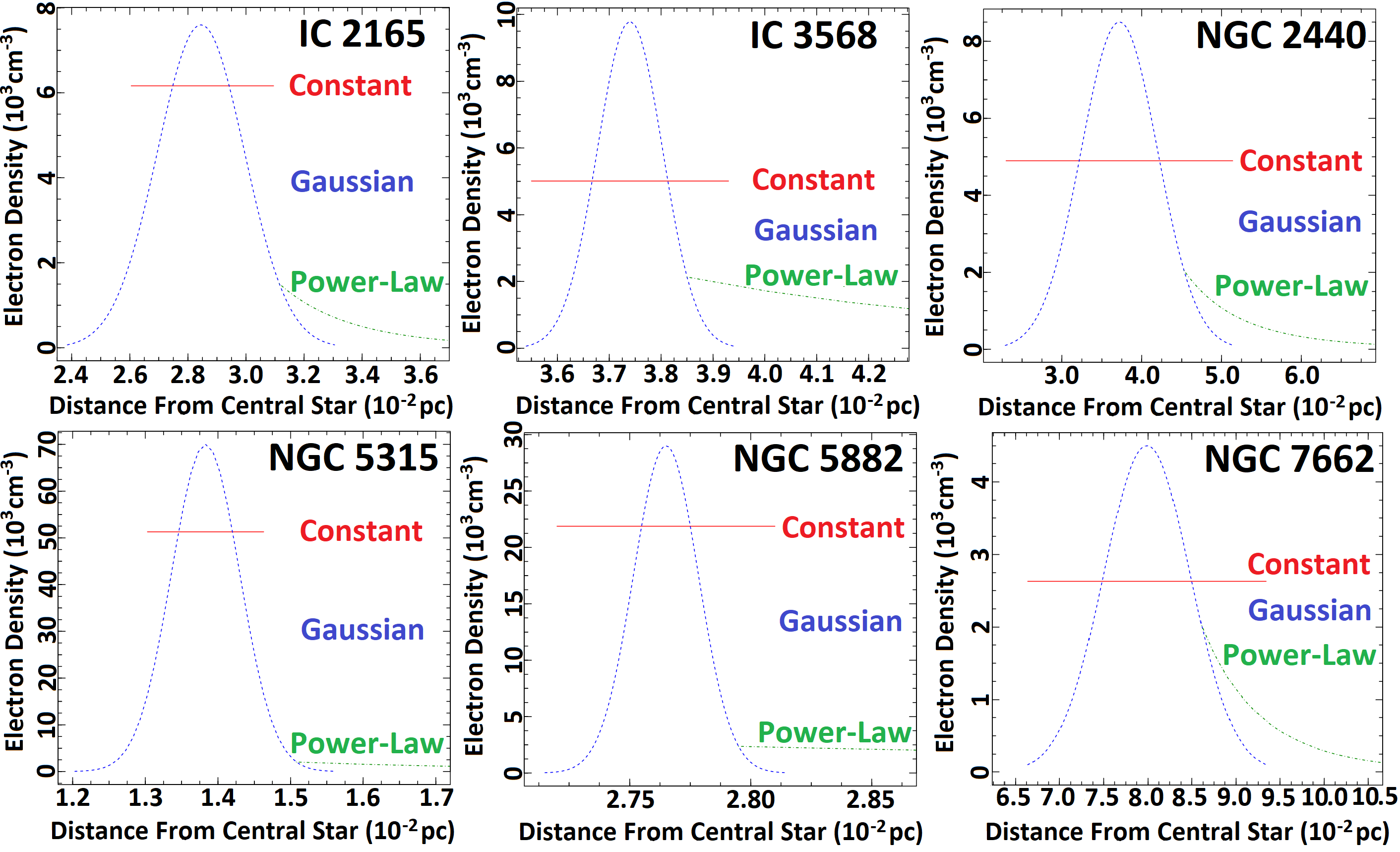}
\label{densfig}
\end{figure*}

In order to constrain the central star's temperature and luminosity for each planetary nebula, Cloudy version 13.03~\citep{fer13} was used to generate matter bounded photoionization models of each PN. For a given set of input parameters, Cloudy simultaneously solves energy and ionization equilibrium equations at specific radial points in the PN model. Three iterations for each model were carried  out to ensure that each model had converged to a solution. The assumed spectral energy distribution (SED) of the central star was taken to be either the Rauch H-Ni or the Rauch pg1159 atmospheric simulations~\citep{rau03}, depending on the known atmospheric properties of the central stars. Rauch H-Ni includes line blanketing for lines from hydrogen to nickel, whereas for Rauch pg1159, the line blanketing effects only include lines of helium, carbon, nitrogen, and oxygen. NGC 5315 was the only PN for which the Rauch pg1159 SED was employed, since its central star was observed to be hydrogen deficient by~\citet{men82}. We assumed a spherically symmetric and static geometry without shock heating of the gas for each model. The initial stellar parameters were chosen to be the average of the most reliable values found in the literature [\citet{sha85}, \citet{sha89}, \citet{zha93}, \citet{cor03}, and \citet{fre08}] but were left variable in the modeling. Fixed angular radii for the models were measured from archival HST WFPC2 images and the values are as follows; IC 2165 r=2.67$''$, IC 3568 r=3.54$''$, NGC 2440 r=6.51$''$, NGC 5315 r=1.50$''$, NGC 5882 r=3.21$''$, and NGC 7662 r=10.71$''$. These fixed angular radii ensured that a change in the outer radius of each model resulted in a change in the assumed distance to the nebula. There are recent GAIA parallax measurements for IC 3568, NGC 5882, and NGC 7662 from \citet{bai18} that were used as initial distance estimates for those planetary nebulae. The inner radius was left completely variable to adjust the thickness of the gas. This allowed for both matter-bounded and radiation-bounded models to be calculated. The best models for each planetary nebula were matter-bounded. Lastly, the nebular properties initially were set to the values discussed in \S3.1 (densities in Table~\ref{ionabun1} and total abundances in Table~\ref{tab:abun1}) but also left variable.
\\\\The calculations for converting the internal model gas properties into model emission line strengths were made with the program PANIC (Paper~III). To summarize the process, radial emissivity values from each Cloudy model are read into PANIC, which then calculates the model line intensities based on the geometry and other properties of the modeled gas. To weight the emission lines equally, only one\footnotemark~emission line per ion was used in the comparison to the models and only for lines having a S/N $>$ 3 (see Tables~\ref{modline1}~\&~\ref{nmodline1} for all lines). A total rms value was calculated for each model from the expression $\sqrt{\frac{1}{N}\sum_{1}^{N}{(1-\frac{model}{observed})^2}}$, where \textit{N} is the total number of lines and diagnostics, \textit{model} is the value of each line or diagnostic predicted by the model, and \textit{observed} is the observed value. The model that yielded the lowest rms value was chosen to be the best model. A similar rms value for the observations was calculated using the line's uncertainty by replacing $1-\frac{model}{observed}$ in the above expression with $\frac{uncertainty}{observed}$. This observed rms value was later used along with the best model rms value to estimate the uncertainties in the parameters from each model. Degeneracies between models with similar rms values were broken by employing five common diagnostics\footnotemark.
\footnotetext[2]{Additional lines from some ions were used in diagnostic line ratios only.}
\footnotetext[3]{([O III] $\lambda$5007+$\lambda$4959+ [O II] $\lambda$3727)/H$\beta$, ([O II] $\lambda$3727)/([O III] $\lambda$5007), (He II $\lambda$4686)/(He I $\lambda$5876), [O III] ($\lambda$4363/$\lambda$5007), and (C III] $\lambda$1909/[C III]$\lambda$1907)}
\\\\To test the sensitivity of the best stellar parameters on the choice of density profiles, a total of three density profiles were used. The first had a constant density throughout the entire gas region (constant density profile). The brightest portions of each nebula are primarily from one shell, so a constant density profile is a sensible first choice. However, the sharp cutoff of the gas at the boundaries is unphysical in nature. Therefore, the second profile replaces the constant density with a Gaussian shaped density profile (Gaussian density profile), allowing for a smoother transition to low density areas. The Gaussian density profile, however, does not take into account the less luminous, exterior shells of gas seen in most of our objects. To represent these outer shells in the third profile, a radially decreasing power-law is appended to the Gaussian density profile (Gaussian with a power-law density profile). Figure~\ref{densfig} shows the different profiles for the full regions of each planetary nebula.
The following process was used to find the best fit parameters for the constant density profile\footnotemark. We began with the initial nebular and stellar parameters described above and proceeded to set other parameters such as grain properties and the filling factor to typical PN values. The density was then adjusted until the carbon density diagnostic, C III] ($\lambda$1909/$\lambda$1907), was within roughly 1\% of the observed value. Next, the radius and overall thickness of the gas was adjusted until the line strength of H$\beta$ was within 10\% of the observed value. The radii of the smaller regions were allowed to vary independent of their full regions' values, while holding the PN distance fixed. In doing so, the spherical symmetry limitation of Cloudy can, to an extent, be compensated for since each smaller region would have its own distance and thickness from the central star. This more accurately depicts the non-spherical PNe as the assymetry guarantees that the gas is not all the same distance from the central star. Finally, all other parameters (abundances, stellar properties, etc.) were adjusted as needed to get the rms down around 0.3, ensuring an average error for all lines and diagnostics near 10\%. However, for the smaller regions, the stellar properties were held fixed at the best solution of the full region.  From there, the best parameters set so far were entered into a script that varied each parameter one at a time, calculated the rms after each model run by calling PANIC, compared the new rms to the previous values, and retained the sets that resulted in a smaller rms. It continuously looped through all parameters until it failed to produce a lower rms value. To minimize convergence times, the step size for each parameter began at relatively large values (e.g. $\bigtriangleup$T=500K, log[L/L$_{\astrosun}$] $\bigtriangleup$dex=0.1, abundance $\bigtriangleup$dex=0.1), but then was reduced to smaller values (e.g. $\bigtriangleup$T=100K, log[L/L$_{\astrosun}$] $\bigtriangleup$dex=0.01, abundance $\bigtriangleup$dex=0.01) to yield the smallest rms and subsequent best parameters.
\footnotetext[4]{This differs from the method employed in Paper~III in that here a large grid of models was not generated first and then compared to each observation. Instead, comparisons were made after each model run to probe only the regions in parameter space that improved the solution.} 
\\\\The Gaussian and Gaussian with a power-law density profiles initially assumed the best fit constant density parameters as the initial values with the inner and outer radii held fixed. Since there is less gas contained within a Gaussian profile of equal height and width relative to a constant profile, the density peak of the Gaussian profile was increased by trial and error to compensate for the decrease in the H$\beta$ line strength. However, for NGC 5315, the increase was found to be too large to keep the carbon density diagnostic within reasonable error, so the inner and outer radii were increased while keeping the peak of the Gaussian distribution at the center of the best constant density model. The slope of the power-law was chosen such that the Gaussian side of the profile was around 2000 cm$^{-3}$ and decreased to about 10 cm$^{-3}$ at the location of the outermost known shell/halo of each planetary nebula. The error estimation process starts with the best model and varies each parameter until the calculated rms is larger than the sum of the best and observed rms values. Only the constant density models have estimated errors since the calculation time in the case of the other two density profiles is prohibitively high.
\\\\Table~\ref{modline1} contains the ratios of the model-predicted strengths to their observed counterparts of important emission lines for each nebular region. The wavelength of each emission line is in the first column followed by the identification of the ion that produces it in the second column. The remaining columns contain the individual region's model$/$observed ratio. The majority of modeled lines for each full region fall within the respective errors shown in Table~\ref{emisstab1}. 
\\\\Table~\ref{nmodline1} provides the same type of comparison as Table~\ref{modline1} but for the other two density profiles and only for the full regions. IC 2165 and NGC 2440 have the largest increase in the number of lines falling outside of the observed errors for the two non-constant density profiles compared to the constant density profile. This suggests that these PN are better represented by the constant density profile.\\\\
\begin{table*}
\centering
\caption{Constant Density Model Emission Lines Compared to Observations: The complete table is available online.}
\label{modline1}
\begin{tabular}{clccccc}
\hline
Wave & & IC2165 Full & IC2165 1 & IC2165 2 & IC2165 3 & IC2165 4 \\ 
\cline{3-7}
(\AA) & ID & Model/Observed & Model/Observed & Model/Observed & Model/Observed & Model/Observed \\
\hline
1485 & N IV] 				& 1.00$\pm$0.11 & 1.00$\pm$0.15 				& 1.00$\pm$0.15 			& 1.00$\pm$0.13 			& 1.01$\pm$0.13 \\     
1907 & [C III]					& 1.01$\pm$0.11 & 0.96$\pm$0.13 				& 1.05$\pm$0.15 			& 1.03$\pm$0.13 			& 1.06$\pm$0.13 \\    
1909\tablenotemark{b} & C III]  		& 1.00$\pm$0.11 & 0.97$\pm$0.13 				& 1.04$\pm$0.15 			& 1.03$\pm$0.13 			& 1.06$\pm$0.13 \\        
3869 & [Ne III]  				& 1.00$\pm$0.07 & 1.00$\pm$0.07 				& 1.00$\pm$0.11 			& 1.00$\pm$0.12 			& 1.01$\pm$0.11 \\  
4363\tablenotemark{b} & [O III]  	& 1.01$\pm$0.11 & 0.87$\pm$0.18	 			& 1.11$\pm$0.14			& 1.04$\pm$0.20 			& 1.10$\pm$0.22 \\       
4686 & He II  				& 1.00$\pm$0.05 & 1.09$\pm$0.06\tablenotemark{a} 	& 0.81$\pm$0.06\tablenotemark{a} & 0.97$\pm$0.06			& 0.85$\pm$0.05\tablenotemark{a} \\
4861 & H$\beta$  				& 0.98$\pm$0.03 & 1.00$\pm$0.04 				& 0.95$\pm$0.05 & 0.94$\pm$0.04\tablenotemark{a} & 0.84$\pm$0.04\tablenotemark{a} \\  
4959\tablenotemark{b} & [O III]  	& 1.03$\pm$0.03 & 0.99$\pm$0.03 				& 1.04$\pm$0.04 & 1.01$\pm$0.03 			& 1.11$\pm$0.03 \\       
5007 & [O III]  				& 1.01$\pm$0.02 & 0.95$\pm$0.03\tablenotemark{a}	& 1.04$\pm$0.03\tablenotemark{a} & 1.03$\pm$0.03 			& 1.08$\pm$0.03\tablenotemark{a} \\   
5876 & He I  					& 1.00$\pm$0.14 & 1.06$\pm$0.28 				& 0.95$\pm$0.41 			& 0.98$\pm$0.32 			& 0.93$\pm$0.37 \\
7136 & [Ar III]  				& 1.00$\pm$0.14 & 1.00$\pm$0.15 				& 1.00$\pm$0.30			& 1.00$\pm$0.15 			& 1.00$\pm$0.22 \\  
9532 & [S III]  				& 1.00$\pm$0.06 & 1.00$\pm$0.12 				& 1.00$\pm$0.10 			& 1.00$\pm$0.12 			& 1.00$\pm$0.14 \\
\hline
\end{tabular}
\tablenotetext{a}{Modeled emission line intensity outside observed error bar.}
\tablenotetext{b}{This line was only included in a diagnostic for the rms calculation for reasons discussed in the text.}
\tablenotetext{c}{Blended 1907 and 1909.}
\end{table*}
\begin{table*}
\centering
\caption{Non-constant Density Model Emission Lines Compared to Observations For Full Regions}
\label{nmodline1}
\begin{tabular}{clcccccc}
\hline
 & & \multicolumn{2}{c}{IC2165 Full} & \multicolumn{2}{c}{IC3568 Full} & \multicolumn{2}{c}{NGC2440 Full}\\
\cline{3-8}
 & & Gaussian & Gaussian With & Gaussian & Gaussian With & Gaussian & Gaussian With\\
Wave & & & power-law & & power-law & & power-law \\ 
\cline{3-8}
(\AA) & ID & Model/Observed & Model/Observed & Model/Observed & Model/Observed & Model/Observed & Model/Observed \\ 
\hline
1485 &  N IV]  				& 1.00$\pm$0.11 			& 1.00$\pm$0.11 				& \nodata 			& \nodata 				& 0.77$\pm$0.12\tablenotemark{a} & 0.77$\pm$0.12\tablenotemark{a}\\
1549 & C IV 					& \nodata 			& \nodata 				& 1.01$\pm$0.13 			& 1.01$\pm$0.13 				& \nodata 			& \nodata \\
1907 &  [C III]  				& 1.04$\pm$0.11 			& 1.06$\pm$0.11 				& 0.99$\pm$0.10 			& 0.99$\pm$0.10 				& 1.02$\pm$0.12 			& 1.00$\pm$0.12  \\  
1909\tablenotemark{b} &  C III]  	& 1.12$\pm$0.11\tablenotemark{a} & 1.14$\pm$0.11\tablenotemark{a} & 1.08$\pm$0.11			& 1.06$\pm$0.11 				& 1.04$\pm$0.12 			& 1.02$\pm$0.12 \\
3727 &  [O II]  				& \nodata 			& \nodata 				& \nodata 			& \nodata 				& 0.83$\pm$0.06\tablenotemark{a} & 0.83$\pm$0.06\tablenotemark{a} \\  
3869 &  [Ne III]  				& 1.00$\pm$0.07 			& 1.00$\pm$0.07 				& 0.99$\pm$0.06 			& 0.99$\pm$0.06 				& 0.99$\pm$0.05 			& 1.02$\pm$0.05\\
4363\tablenotemark{b} &  [O III]  	& 1.13$\pm$0.11\tablenotemark{a} & 1.15$\pm$0.11\tablenotemark{a} & 1.05$\pm$0.31			& 0.99$\pm$0.31 				& 1.06$\pm$0.10 			& 1.04$\pm$0.10 \\ 
4686 &  He II  				& 0.99$\pm$0.05 			& 1.00$\pm$0.05 				& \nodata 			& \nodata 				& 1.12$\pm$0.04\tablenotemark{a} & 1.10$\pm$0.04\tablenotemark{a} \\
4861 &  H$\beta$  				& 0.97$\pm$0.03			 & 0.98$\pm$0.03 			& 0.99$\pm$0.03 			& 1.02$\pm$0.03 				& 1.03$\pm$0.04 			& 1.02$\pm$0.04\\
4959\tablenotemark{b} &  [O III]  	& 1.08$\pm$0.03\tablenotemark{a} & 1.09$\pm$0.03\tablenotemark{a} & 1.19$\pm$0.02\tablenotemark{a} & 1.12$\pm$0.02\tablenotemark{a} & 1.10$\pm$0.02\tablenotemark{a} & 1.07$\pm$0.02\tablenotemark{a} \\ 
5007 &  [O III]  				& 1.06$\pm$0.02\tablenotemark{a} & 1.07$\pm$0.02\tablenotemark{a} & 1.06$\pm$0.02\tablenotemark{a} & 1.00$\pm$0.02 				& 1.08$\pm$0.02\tablenotemark{a} & 1.06$\pm$0.02\tablenotemark{a}\\ 
5876 &  He I  				& 0.98$\pm$0.14 			& 0.99$\pm$0.14 				& 0.99$\pm$0.14 			& 1.00$\pm$0.14 				& 1.03$\pm$0.10 			& 1.02$\pm$0.10 \\  
6584 &  [N II]				& \nodata 			& \nodata 				& \nodata 			& \nodata 				& 1.16$\pm$0.00\tablenotemark{a} & 1.16$\pm$0.00\tablenotemark{a} \\
7136 &  [Ar III]  				& 1.00$\pm$0.14 			& 0.99$\pm$0.14 				& 0.99$\pm$0.21 			& 1.00$\pm$0.21 				& 1.08$\pm$0.04\tablenotemark{a} & 1.07$\pm$0.04\tablenotemark{a} \\
9532 &  [S III]  				& 1.00$\pm$0.06 			& 0.99$\pm$0.06 				& \nodata 			& \nodata 				& 1.05$\pm$0.04\tablenotemark{a} & 1.04$\pm$0.04 \\ 
\hline
 & & \multicolumn{2}{c}{NGC5315 Full} & \multicolumn{2}{c}{NGC5882 Full} & \multicolumn{2}{c}{NGC7662 Full}\\
\cline{3-8}
 & & Gaussian & Gaussian With & Gaussian & Gaussian With & Gaussian & Gaussian With\\
  & & & power-law & & power-law & & power-law \\ 
\cline{3-8}
 & & Model/Observed & Model/Observed & Model/Observed & Model/Observed & Model/Observed & Model/Observed \\ 
\hline
1485 &  N IV] 					& \nodata 					& \nodata 					& \nodata 					& \nodata 					& 1.02$\pm$0.06 				& 1.11$\pm$0.06\tablenotemark{a} \\
1549 & C IV 						& 1.00$\pm$0.04 				& 0.99$\pm$0.04 				& 1.00$\pm$0.24 				& 1.16$\pm$0.24 				& 0.93$\pm$0.04\tablenotemark{a} 	& 1.02$\pm$0.04 \\
1750 & N III] 					& \nodata 					& \nodata 					& \nodata 					& \nodata 					& 0.98$\pm$0.24 				& 0.97$\pm$0.23 \\
1907 &  [C III]  					& 0.99$\pm$0.07 				& 0.99$\pm$0.07 				& 1.00$\pm$0.14 				& 1.00$\pm$0.14 				& 1.03$\pm$0.04 				& 1.02$\pm$0.04 \\
1909\tablenotemark{b} &  C III]  		& 1.07$\pm$0.05\tablenotemark{a}	& 1.07$\pm$0.05\tablenotemark{a} 	& 1.00$\pm$0.11 				& 0.99$\pm$0.11 				& 1.06$\pm$0.05\tablenotemark{a} 	& 1.02$\pm$0.05 \\
3727 & [O II] 					& \nodata 					& \nodata 					& \nodata 					& \nodata 					& 0.92$\pm$0.30 				& 1.04$\pm$0.35 \\
3869 &  [Ne III]  					& 1.00$\pm$0.01 				& 0.99$\pm$0.01 				& 1.03$\pm$0.04 				& 1.07$\pm$0.04\tablenotemark{a} 	& 1.00$\pm$0.03 				& 1.03$\pm$0.03\tablenotemark{a} \\
4363\tablenotemark{b} &  [O III]  		& 1.05$\pm$0.06 				& 1.02$\pm$0.06 				& 1.12$\pm$0.20 				& 1.14$\pm$0.20 				& 0.97$\pm$0.06			 	& 0.88$\pm$0.06\tablenotemark{a} \\
4686 &  He II  					& \nodata 					& \nodata 					& 0.99$\pm$0.24 				& 0.96$\pm$0.24 				& 1.03$\pm$0.02\tablenotemark{a} 	& 1.08$\pm$0.02\tablenotemark{a} \\
4861 &  H$\beta$  					& 1.03$\pm$0.01\tablenotemark{a} 	& 1.04$\pm$0.01\tablenotemark{a} 	& 0.81$\pm$0.02\tablenotemark{a} 	& 0.86$\pm$0.02\tablenotemark{a} 	& 1.01$\pm$0.01			 	& 1.04$\pm$0.01\tablenotemark{a} \\
4959\tablenotemark{b} &  [O III]  		& 1.01$\pm$0.01 				& 1.00$\pm$0.01 				& 1.32$\pm$0.02\tablenotemark{a} 	& 1.37$\pm$0.02\tablenotemark{a} 	& 1.10$\pm$0.01\tablenotemark{a} 	& 1.06$\pm$0.01\tablenotemark{a}\\
5007 &  [O III]  					& 0.99$\pm$0.00\tablenotemark{a} 	& 0.98$\pm$0.00\tablenotemark{a}	& 1.11$\pm$0.02\tablenotemark{a}	& 1.16$\pm$0.02\tablenotemark{a} 	& 1.00$\pm$0.01			 	& 0.97$\pm$0.01\tablenotemark{a} \\
5876 &  He I  					& 1.01$\pm$0.01 				& 1.00$\pm$0.01 				& 0.99$\pm$0.05 				& 1.05$\pm$0.05 				& 1.03$\pm$0.14 				& 1.06$\pm$0.15 \\
6584 & [N II] 					& \nodata 					& \nodata 					& 1.00$\pm$0.03 				& 0.98$\pm$0.03 				& \nodata 					& \nodata \\
7136 &  [Ar III]  					& 0.98$\pm$0.00\tablenotemark{a} 	& 1.00$\pm$0.00 				& 1.00$\pm$0.05 				& 0.99$\pm$0.05 				& 1.00$\pm$0.12 				& 1.04$\pm$0.13 \\
9532 &  [S III]  					& 0.95$\pm$0.01\tablenotemark{a} 	& 0.95$\pm$0.01\tablenotemark{a} 	& 1.01$\pm$0.03 				& 1.01$\pm$0.03 				& 1.00$\pm$0.09 				& 1.03$\pm$0.09 \\
\hline
\end{tabular}
\tablenotetext{a}{Modeled emission line intensity outside observed error bar.}
\tablenotetext{b}{This line was only included in a diagnostic for the rms calculation for reasons discussed in the text.}
\end{table*}
Table~\ref{condenmods1} contains the best parameter values and associated errors for the constant density models. Names of the stellar and nebular parameters that were used in the modeling are in the first column along with the values for each region in the subsequent columns. The model abundances are also shown alongside the observed abundances in Figures~\ref{fig:abun1}-\ref{fig:abun3}. Only the models for NGC 5315 required abundances significantly different from the observed values, especially for sulphur and argon. After the modeling parameters, we present the nebular temperatures, densities, ionization correction factors, and rms values for each region. The model temperatures and densities were calculated by running the model emission lines (from PANIC) through ELSA. For the model ICFs, they were calculated by taking each model abundance and dividing it by the sum of the necessary ionic model abundances (as determined by ELSA from the PANIC emission lines) for a direct comparison to the ICFs in Table~\ref{ionabun1}. For example, the model ICF for oxygen was calculated by dividing the model oxygen abundance by the sum of the model ionic abundances for O$^+$ and O$^{+2}$ since that ratio is what the oxygen ICFs in Table~\ref{ionabun1} are supposed to represent. All but the rms values are scaled with respect to the observed counterpart. For most of the regions, the model rms is below the observed rms, implying a good fit. The smallest model rms of 0.0063 from the full region of IC 2165 shows that the assumed model represented that region's structure quite well. In most cases, the modeled temperatures and densities of the gas are within the observed errors. Also, the model abundances for the regions agree within error with the observed abundances in most cases. As previously noted in Paper~III, the asymmetry in the errors is the result of the rms equation having a lower bound.\\\\
\begin{table*}
\centering
\caption{Constant Density Models: The complete table is available online.}
\label{condenmods1}
\begin{tabular}{cccccc}
\hline
Parameter & IC2165 Full & IC2165 1 & IC2165 2 & IC2165 3 & IC2165 4 \\
\hline
$T_{star}$ (kK) & 110.0$^{+7.0}_{-17.0}$ & 110.0 & 110.0 & 110.0& 110.0  \\
$L_{star}$ (log[L/$L_{\astrosun}$]) & 3.17$^{+0.20}_{-0.14}$ & 3.17 & 3.17 & 3.17 & 3.17  \\
Distance (kpc) & 2.39$^{+0.36}_{-0.33}$ & 2.39 & 2.39 & 2.39 & 2.39 \\
$H_{den}$ (log[$H_{density}$]) & 3.79$^{+0.02}_{-0.03}$ & 3.75$^{+0.04}_{-0.04}$ & 4.06$^{+0.05}_{-0.07}$ & 3.70$^{+0.04}_{-0.04}$ & 3.75$^{+0.04}_{-0.05}$ \\
Inner Radius ($10^{-2}$pc) & 2.60$^{+0.06}_{-0.06}$ & 2.56$^{+0.11}_{-0.15}$ & 2.60$^{+0.05}_{-0.04}$ & 2.38$^{+0.15}_{-0.16}$ & 2.47$^{+0.13}_{-0.13}$ \\
Outer Radius ($10^{-2}$pc) & 3.09$^{+0.05}_{-0.05}$ & 3.09$^{+0.08}_{-0.07}$ & 2.73$^{+0.04}_{-0.05}$ & 3.11$^{+0.12}_{-0.11}$ & 2.96$^{+0.12}_{-0.12}$ \\
Filling Factor ($10^{-1}$) & 9.89$^{+0.11}_{-0.99}$ & 10.00$^{+0.00}_{-0.16}$ & 9.49$^{+0.51}_{-2.97}$ & 10.00$^{+0.00}_{-0.17}$ & 10.0$^{+0.00}_{-0.25}$ \\
He/H ($10^{-2}$) & 10.28$^{+1.41}_{-1.29}$ & 11.07$^{+2.58}_{-2.03}$ & 10.26$^{+4.20}_{-5.03}$ & 10.33$^{+2.55}_{-2.01}$ & 10.47$^{+3.33}_{-3.23}$ \\
C/H ($10^{-4}$) & 4.57$^{+2.41}_{-1.59}$ & 4.75$^{+5.55}_{-2.63}$ & 2.86$^{+5.15}_{-2.16}$ & 4.35$^{+4.78}_{-2.30}$ & 3.80$^{+5.97}_{-2.45}$ \\
N/H ($10^{-4}$) & 1.10$^{+0.43}_{-0.43}$ & 0.78$^{+0.56}_{-0.55}$ & 1.38$^{+1.52}_{-1.35}$ & 0.94$^{+0.34}_{-0.66}$ & 0.95$^{+0.91}_{-0.87}$ \\
O/H ($10^{-4}$) & 2.69$^{+0.91}_{-0.80}$ & 2.89$^{+1.95}_{-1.49}$ & 1.75$^{+1.80}_{-1.38}$ & 2.68$^{+1.89}_{-1.42}$ & 2.82$^{+2.43}_{-1.84}$ \\
Ne/H ($10^{-5}$) & 3.39$^{+1.32}_{-1.31}$ & 3.70$^{+2.70}_{-2.63}$ & 2.10$^{+2.14}_{-2.06}$ & 2.98$^{+2.15}_{-2.07}$ & 2.90$^{+2.72}_{-2.65}$ \\
S/H ($10^{-6}$) & 3.73$^{+1.42}_{-1.44}$ & 3.56$^{+2.53}_{-2.50}$ & 2.15$^{+2.11}_{-2.11}$ & 5.19$^{+3.72}_{-3.57}$ & 4.44$^{+4.08}_{-4.02}$ \\
Ar/H ($10^{-7}$) & 20.09$^{+7.58}_{-7.76}$ & 24.89$^{+17.77}_{-17.48}$ & 9.86$^{+9.86}_{-9.68}$ & 26.06$^{+17.6}_{-18.12}$ & 24.66$^{+22.65}_{-22.47}$ \\
\hline
[O III] $T^{mod}_{e}$/$T^{obs}_{e}$ & 0.99$\pm$0.04 & 1.00$\pm$0.07 & 1.03$\pm$0.06 & 1.00$\pm$0.08 & 1.01$\pm$0.09 \\
$ $[S II] $N_{e}^{mod}$/$N_{e}^{obs}$ & 5.27$\pm$13.34 & \nodata & \nodata & \nodata & \nodata \\
C III] $N_{e}^{mod}$/$N_{e}^{obs}$ & 0.94$\pm$0.15 & 1.00$\pm$0.27 & 0.99$\pm$0.16 & 1.07$\pm$0.36 & 0.99$\pm$0.23 \\
icf (O)$^{mod}$/icf (O)$^{obs}$  & 0.78$\pm$0.11 & 0.74$\pm$0.11 & 0.52$\pm$0.14 & 0.69$\pm$0.14 & 0.55$\pm$0.15 \\
icf (O)$^{mod}$/icf (O)$^{DMS}$ & 0.91$\pm$0.13 & 0.93$\pm$0.13 & 0.74$\pm$0.10 & 0.91$\pm$0.12 & 0.79$\pm$0.09 \\
icf (Ar)$^{mod}$/icf (Ar)$^{obs}$  & 0.66$\pm$0.09 & 0.66$\pm$0.10 & 0.68$\pm$0.18 & 0.60$\pm$0.12 & 0.31$\pm$0.08 \\
icf (Ne)$^{mod}$/icf (Ne)$^{obs}$  & 1.85$\pm$0.31 & 1.84$\pm$0.29 & 1.74$\pm$0.48 & 2.29$\pm$0.48 & 4.55$\pm$1.22 \\
icf (S)$^{mod}$/icf (S)$^{obs}$  & 2.33$\pm$0.40 & 2.40$\pm$0.42 & 1.90$\pm$0.27 & 2.65$\pm$0.77 & 1.31$\pm$0.87 \\
Model RMS ($10^{-2}$)  & 0.63 & 4.00 & 7.00 & 2.13 & 6.86 \\
Observed RMS ($10^{-2}$)  & 9.35 & 14.53 & 19.35 & 15.8 & 17.69\\
\hline
\end{tabular}
\tablenotetext{a}{Model Returned High Density Limit Default: 20000 cm$^{-3}$.}
\tablenotetext{DMS}{Using the formalism of \citet{del14}.}
\end{table*}
As can be seen in Table~\ref{condenmods1}, the oxygen and argon model ionization correction factors for IC 3568, NGC 5315, and NGC 5882 are larger than their observed counterparts, while the same ICFs are smaller for IC 2165, NGC 2440, and NGC 7662. Also, IC 3568, NGC 5315, and NGC 5882 have lower stellar temperatures than IC 2165, NGC 2440, and NGC 7662. Therefore, the ICF equations employed by ELSA could be under-correcting for PN with lower stellar temperatures and over-correcting for those with higher stellar temperatures. Both of the ICF equations for argon and oxygen depend on the ratio (He$^{+}$+He$^{+2}$)/He$^{+}$. For lower stellar temperatures, the He$^{+}$ ionic state will be populated more than in the case of a higher stellar temperature, reducing the ratio and subsequent ICF value. A recent paper in the literature, \citet{del14}, formulated new ionization correction factors for all of the elements presented here. Given their validity range restrictions, however, for argon and oxygen, only the oxygen ICF is applicable to compare to our work. As can be seen in Table~\ref{condenmods1}, in most cases their oxygen ICF is closer to the model value.
\\\\The neon model ICFs are smaller in all PNe except for NGC 5315 and IC 2165, and upon inspection of the ICF equation, a similar cause like in the cases of the oxygen and argon ICFs above can not be found to explain the differences. Calculating the average observed Ne/O for our sample and comparing to the average modeled Ne/O shows a discrepancy of about 10\%, 0.20$\pm$0.02 and 0.18 for the observed and modeled ratio, respectively. Thus, on average, the systematic error introduced by using ICFs seems to be minor in this case. Interestingly for neon, many regions have model ICFs less than 1, which points to a problem in the neon ionic abundance calculations made by ELSA. This could indicate that the atomic constants such as the collision strengths or Einstein A values are slightly off from their true values, resulting in an ionic abundance of neon that is too large. Lastly, all PN except NGC 5882 have model ICFs of sulphur larger than their observed ICFs with the largest discrepancy seen in NGC 5315. 
\\\\The parameters for the best fit models for the Gaussian and Gaussian with a power-law density profiles are in Table~\ref{gaudenmods1}. This table has the same format as Table~\ref{condenmods1} but only contains the full regions. The difference between these stellar parameters and the constant density models is small in most cases. Only IC 3568 had a somewhat large change in the luminosity of 0.16 dex. However, this is small compared to the errors estimated for the best fit constant density model. Therefore, as stated in Paper~III and further reinforced in this work, the choice of density structure only has a minor effect on the final stellar temperature and luminosity and the adoption of the stellar parameters from the constant density models is reasonable. The Gaussian and Gaussian with a power-law density profiles also generate similar electron temperatures but higher electron densities compared with the constant density profile. This makes sense, since these two profiles have peak density values significantly above the constant density profile (see Figure~\ref{densfig}). The ionization correction factors are either slightly larger or smaller than the constant density ICFs as well. This could be an indication that the choice of density structure affects the ICFs. Lastly, the rms values for IC 2165 and NGC 2440 are considerably larger than their constant density counterparts while the Gaussian profile value for NGC 5882 is smaller than its constant density rms value. This indicates that IC 2165 and NGC 2440 are better represented using a constant density model while a Gaussian density profile appears more suitable in the case of NGC 5882.
\begin{table*}
\centering
\caption{Non-constant Density Models}
\label{gaudenmods1}
\begin{tabular}{ccccccc}
\hline
 & \multicolumn{2}{c}{IC2165 Full} & \multicolumn{2}{c}{IC3568 Full} & \multicolumn{2}{c}{NGC2440 Full}\\
\cline{2-7}
 & Gaussian & Gaussian With & Gaussian & Gaussian With & Gaussian & Gaussian With\\
Parameter & & power-law & & power-law & & power-law \\ 
\hline
$T_{star}$ (kK) & 110.0 & 110.0 & 69.2 & 66.6 & 171.8 & 171.9 \\
$L_{star}$ (log[L/$L_{\astrosun}$]) & 3.19 & 3.19 & 3.67 & 3.68 & 2.74 & 2.73\\
Filling Factor ($10^{-1}$) & 9.51 & 9.45 & 9.70 & 8.67 & 1.76 & 1.71\\
He/H ($10^{-2}$) & 10.19 & 10.14 & 9.27 & 9.04 & 12.27 & 12.27\\
C/H ($10^{-4}$) & 3.84 & 3.92 & 5.35 & 5.24 & 2.54 & 2.54\\
N/H ($10^{-4}$) & 1.18 & 1.15 & 0.15 & 0.17 & 0.74 & 0.74\\
O/H ($10^{-4}$) & 2.54 & 2.54 & 4.83 & 4.45 & 2.71 & 2.69\\
Ne/H ($10^{-5}$) & 3.04 & 3.00 & 7.69 & 7.57 & 3.42 & 3.58\\
S/H ($10^{-6}$) & 2.80 & 2.82 & 12.08 & 13.30 & 1.49 & 1.49\\
Ar/H ($10^{-7}$) & 14.26 & 14.35 & 15.35 & 16.11 & 13.03 & 13.00\\
\hline
[O III] $T^{mod}_{e}$/$T^{obs}_{e}$ & 1.01$\pm$0.04 & 1.01$\pm$0.04 & 0.99$\pm$0.09 & 0.99$\pm$0.09 & 0.99$\pm$0.04 & 0.99$\pm$0.04\\
$ $[S II] $N_{e}^{mod}$/$N_{e}^{obs}$ & 8.4$\pm$21.28 & 8.33$\pm$21.11 & \nodata & \nodata & 5.12$\pm$5.12 & 5.00$\pm$5.00\\
C III] $N_{e}^{mod}$/$N_{e}^{obs}$ & 1.53$\pm$0.24 & 1.52$\pm$0.24 & 1.72$\pm$1.00 & 1.60$\pm$0.94 & 1.17$\pm$0.08 & 1.14$\pm$0.08\\
icf (O)$^{mod}$/icf (O)$^{obs}$ & 0.73$\pm$0.11 & 0.73$\pm$0.11 & 1.07$\pm$0.01 & 1.06$\pm$0.01 & 0.59$\pm$0.05 & 0.58$\pm$0.05\\
icf (O)$^{mod}$/icf (O)$^{DMS}$ & 0.85$\pm$0.12 & 0.85$\pm$0.12 & 1.08$\pm$0.04 & 1.07$\pm$0.04 & 0.69$\pm$0.09 & 0.69$\pm$0.09 \\
icf (Ar)$^{mod}$/icf (Ar)$^{obs}$ & 0.61$\pm$0.08 & 0.61$\pm$0.08 & 0.68\tablenotemark{b} & 0.71\tablenotemark{b} & 0.45$\pm$0.04 & 0.45$\pm$0.04\\
icf (Ne)$^{mod}$/icf (Ne)$^{obs}$ & 0.55$\pm$0.09 & 0.55$\pm$0.09 & 0.91$\pm$0.02 & 0.92$\pm$0.02 & 0.45$\pm$0.05 & 0.46$\pm$0.05\\
icf (S)$^{mod}$/icf (S)$^{obs}$ & 1.67$\pm$0.29 & 1.71$\pm$0.29 & \nodata & \nodata & 1.05$\pm$0.08 & 1.04$\pm$0.08\\
RMS ($10^{-2}$) & 3.40 & 3.60 & 3.13 & 2.31 & 10.25 & 9.97 \\
\hline
 & \multicolumn{2}{c}{NGC5315 Full} & \multicolumn{2}{c}{NGC5882 Full} & \multicolumn{2}{c}{NGC7662 Full}\\
\cline{2-7}
 & Gaussian & Gaussian With & Gaussian & Gaussian With & Gaussian & Gaussian With\\
 & & power-law & & power-law & & power-law \\
\hline
$T_{star}$ (kK) & 69.9 & 69.9 & 79.2 & 78.7 & 109.1 & 109.9 \\
$L_{star}$ (log[L/$L_{\astrosun}$]) & 4.92 & 4.94 & 3.42 & 3.42 & 3.88 & 3.87\\
Filling Factor ($10^{-1}$) & 9.68 & 9.75 & 8.30 & 8.26 & 7.82 & 7.66\\
He/H ($10^{-2}$) & 11.94 & 11.80 & 12.62 & 12.59 & 9.42 & 9.68\\
C/H ($10^{-4}$) & 16.87 & 17.82 & 4.29 & 4.29 & 4.95 & 5.08\\
N/H ($10^{-4}$) & 0.57 & 0.57 & 2.41 & 0.24 & 0.63 & 0.62\\
O/H ($10^{-4}$) & 6.59 & 6.52 & 12.88 & 12.88 & 3.71 & 3.78\\
Ne/H ($10^{-5}$) & 14.89 & 14.83 & 23.66 & 23.66 & 5.81 & 5.86\\
S/H ($10^{-6}$) & 431.52 & 454.99 & 7.66 & 7.69 & 3.19 & 3.18\\
Ar/H ($10^{-7}$) & 959.40 & 1016.25 & 38.11 & 38.46 & 23.23 & 23.33\\
\hline
[O III] $T^{mod}_{e}$/$T^{obs}_{e}$ & 1.00$\pm$0.02 & 1.00$\pm$0.02 & 0.99$\pm$0.05 & 0.98$\pm$0.05 & 0.99$\pm$0.02 & 0.97$\pm$0.02\\
$ $[S II] $N_{e}^{mod}$/$N_{e}^{obs}$ & 1.85$\pm$0.81\tablenotemark{a} & 1.85$\pm$0.81\tablenotemark{a} & 14.29$\pm$28.57\tablenotemark{a} & 14.29$\pm$28.57\tablenotemark{a} & \nodata & \nodata \\
C III] $N_{e}^{mod}$/$N_{e}^{obs}$ & 1.13$\pm$0.14 & 1.14$\pm$0.14 & 0.99$\pm$0.33 & 0.96$\pm$0.32 & 1.32$\pm$0.50 & 0.86$\pm$0.32\\
icf (O)$^{mod}$/icf (O)$^{obs}$ & 1.39\tablenotemark{b} & 1.40\tablenotemark{b} & 1.09$\pm$0.01 & 1.09$\pm$0.01 & 0.78$\pm$0.06 & 0.78$\pm$0.06\\
icf (O)$^{mod}$/icf (O)$^{DMS}$ & 1.36$\pm$0.07 & 1.37$\pm$0.07 & 1.10$\pm$0.04 & 1.10$\pm$0.04 & 0.97$\pm$0.14 & 0.98$\pm$0.14 \\
icf (Ar)$^{mod}$/icf (Ar)$^{obs}$ & 1.94$\pm$0.02 & 1.95$\pm$0.02 & 1.13$\pm$0.01 & 1.17$\pm$0.01 & 0.69$\pm$0.05 & 0.64$\pm$0.05\\
icf (Ne)$^{mod}$/icf (Ne)$^{obs}$ & 1.21$\pm$0.01 & 1.22$\pm$0.01 & 0.95$\pm$0.01 & 0.95$\pm$0.01 & 0.61$\pm$0.04 & 0.56$\pm$0.04\\
icf (S)$^{mod}$/icf (S)$^{obs}$ & 23.55$\pm$4.08 & 25.19$\pm$4.36 & 0.49$\pm$0.15 & 0.53$\pm$0.16 & 1.21$\pm$0.28 & 1.14$\pm$0.27\\
RMS ($10^{-2}$) & 3.32 & 3.45 & 5.52 & 7.97 & 4.18 & 4.69 \\
\hline
\end{tabular}
\tablenotetext{a}{Model Returned High Density Limit Default: 20000 cm$^{-3}$.}
\tablenotetext{DMS}{Using the formalism of \citet{del14}.}
\tablenotetext{b}{\footnotesize Error unknown since observed value is assumed.}
\end{table*}

\section{Discussion}

\begin{figure}
\caption{Comparison of the best fit constant density stellar temperature and luminosity for each planetary nebula to other values in the literature.\newline}
\includegraphics[scale=0.35]{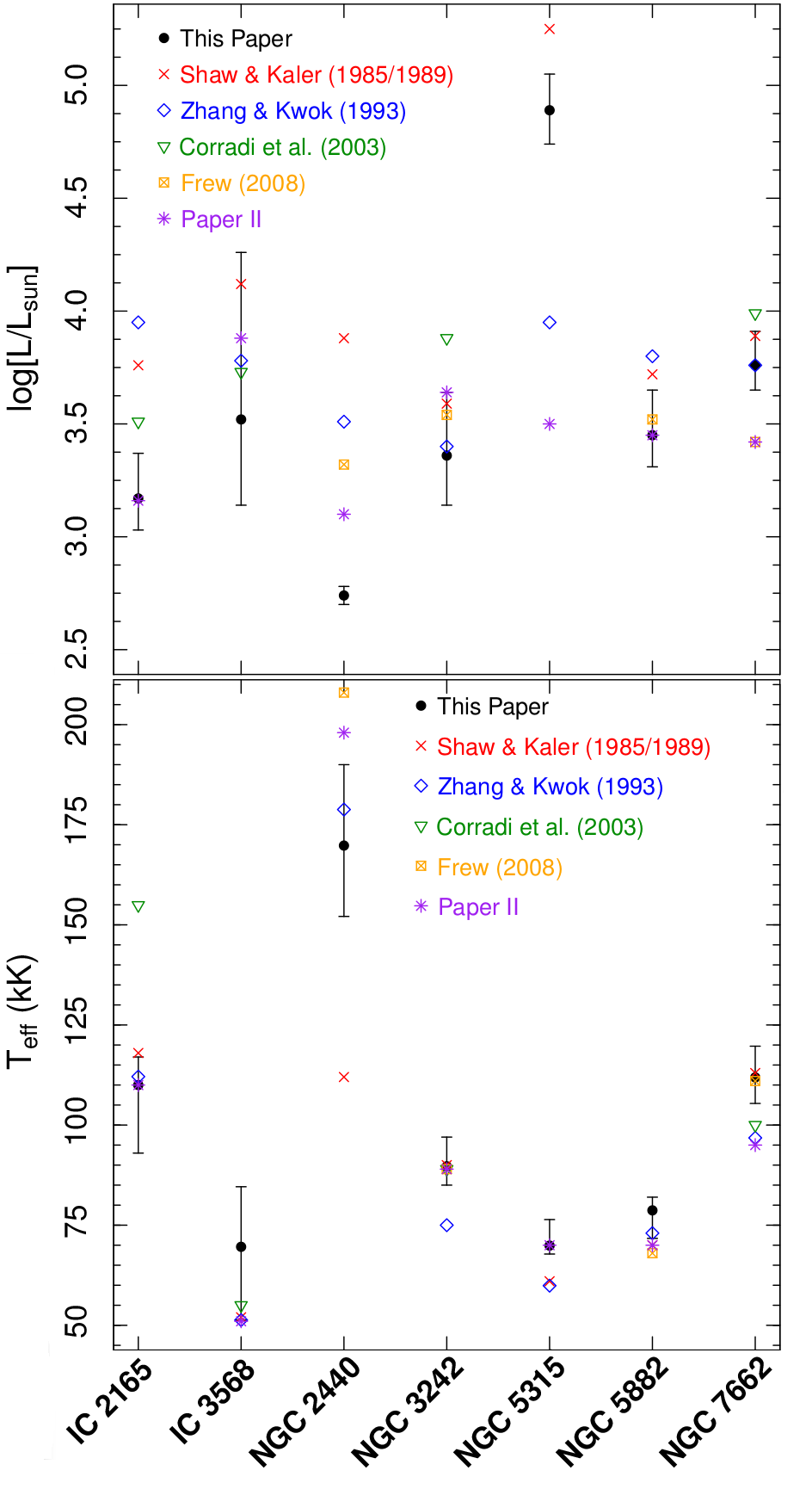}
\label{tefflumfig}
\end{figure}

\begin{figure}
\caption{Log \textit{L/L$_{sun}$} vs. log \textit{T$_{eff}$} for the entire planetary nebula sample, including NGC 3242 from Paper~III. Post-AGB model tracks from \citet{vas94} (VW 1994, solid red lines, Z=0.016), \citet{sch83} (S 1983, blue dashed lines, Z=0.016) and \citet{ber16} (MB 2016, purple dash-dot lines, Z=0.02 and green dash lines, Z=0.01) are overlaid. Beneath each author on the plot is the ZAMS/final mass in solar masses following each evolutionary track from top to bottom.}
\includegraphics[scale=0.37]{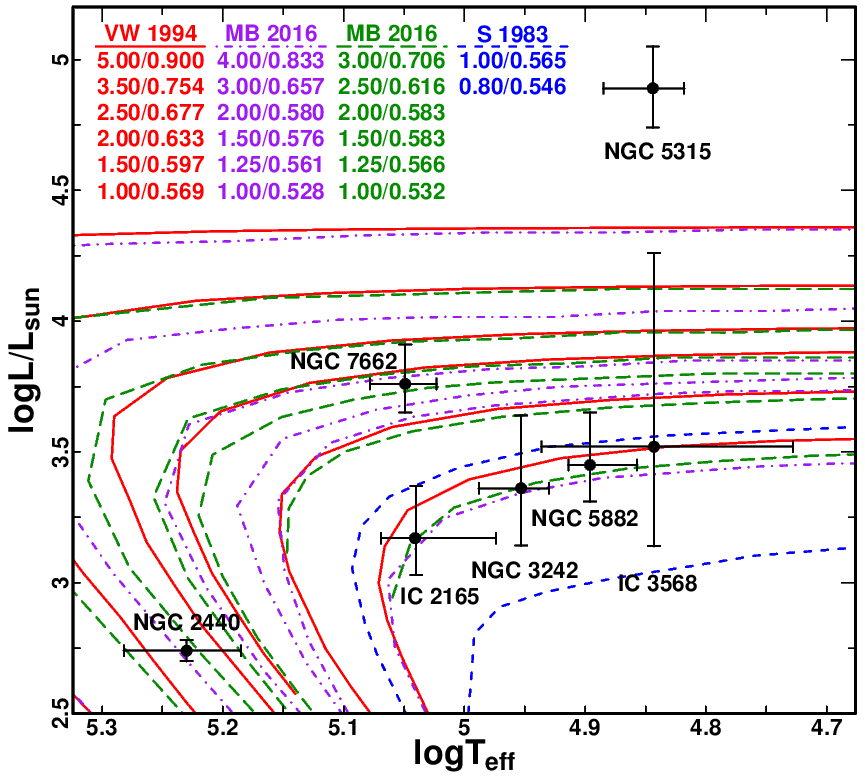}
\label{agbmods}
\end{figure}

One of the major results of our study is that within our uncertainties, the elements He, C, N, O, Ne, Ar, and S are distributed homogeneously across each planetary nebula in our sample (Figures~\ref{fig:abun1}-\ref{fig:abun3} and Table~\ref{tab:abun1}). This is the first time these planetary nebulae have been shown to be chemically homogeneous at sub-arcsecond, effective spatial resolution. This abundance homogeneity for each PN supports current planetary nebula formation theory. During the asymptotic giant branch stage of a star's evolution, convection processes will mix recent nuclear fusion products like carbon, nitrogen, and helium into the surface layers which are eventually ejected during the formation of the PN. The relevant timescales are those of nuclear fusion and envelope convection. The nuclear timescale is of the order 10$^5$-10$^6$ years whereas the convection process is model dependent and the timescale is calculated by the taking the size of the envelope divided by the velocity of the convective atmospheric elements \citep{ren81}. Models by \citet{bue97} (private communication) imply a convection timescale on the order of a year. Therefore, the mixing and subsequent ejection of matter to form the planetary nebula is much shorter than the nuclear timescales, and each ejection should be homogeneous. Our results here confirm that indeed the convection timescale must be much shorter than the nuclear timescale. Also, if the shells observed in IC 2165, IC 3568, and NGC 5882 are from different ejection events, then the ejection timescale must also be shorter than the nuclear timescale since the abundances between shells were the same.
\\\\We compare the stellar parameters determined by various authors over the last three decades to the best-fit constant density values in Table~\ref{tlstar} and Figure~\ref{tefflumfig}. The first column in Table~\ref{tlstar} gives the name of each sample PN followed in subsequent columns by the luminosities [in log(L/L\textsubscript{\astrosun})] and temperatures (in kK) for each source. As can be seen in Figure~\ref{tefflumfig}, our stellar parameters for IC 2165 agree within the 1$\sigma$ uncertainties with those from Paper~II and our temperature agrees with that of \citet{zha93}. The stellar parameters for IC 3568 are consistent with the other authors, except for the temperature of Paper~II (300 K smaller than our lower limit). Our luminosity for NGC 2440 is significantly lower than the other published values, although the temperature matches (given the uncertainties) with that of \citet{zha93}. On the other hand, the opposite is seen for NGC 5315 where the luminosity is significantly higher compared to the values from the other authors, except for the upper bound determined by \citet{sha89}. The temperatures for NGC 5315 and NGC 5882 are consistent with Paper~II and \citet{zha93}, respectively. The luminosity for NGC 5882 matches with the values of \citet{fre08} and Paper~II, within uncertainties, and NGC 7662's luminosity is consistent with those of \citet{zha93} and \citet{sha89}.  Lastly, \citet{sha85} and \citet{fre08} have temperatures consistent with our value for NGC 7662. \\\\
\begin{table*}
\centering
\caption{Comparison: Stellar Parameters}
\label{tlstar}
\begin{tabular}{lcccccc}
\hline
Object & Model & SK85/89\tablenotemark{a} & ZK93\tablenotemark{b} & CSSP03\tablenotemark{c} & F08\tablenotemark{d} & Paper~II\tablenotemark{e}\\
\cline{1-7}
 & \multicolumn{5}{c}{Luminosity (log[L/L\textsubscript{\astrosun}])}\\
 IC 2165 & 3.17 & 3.76 & 3.95 & 3.51 & \nodata & 3.16\\
IC 3568 & 3.52 & 4.12 & 3.78 & 3.73 & \nodata & 3.88\\
NGC 2440 & 2.74 & 3.88 & 3.51 & \nodata & 3.32 & 3.10\\
NGC 5315 & 4.89 & $<$5.25 & 3.95 & \nodata & \nodata & 3.50\\
NGC 5882 & 3.45 & 3.72 & 3.80 & \nodata & 3.52 & 3.45\\
NGC 7662 & 3.76 & 3.89 & 3.76 & 3.99 & 3.42 & 3.42\\
\hline
 & \multicolumn{5}{c}{Effective Temperature (kK)}\\
\hline
IC 2165 & 110.0 & 118.0 & 112.1 & 154.9 & \nodata & 110.0\\
IC 3568 & 69.6 & 52.0 & 51.3 & 55.0 & \nodata & 51.0\\
NGC 2440 & 169.8 & 112.0 & 178.8 & \nodata & 208.0 & 198.0\\
NGC 5315 & 69.9 & 61.0 & 59.9 & \nodata & \nodata & 70.0\\
NGC 5882 & 78.7 & 70.0 & 73.0 & \nodata & 68.0 & 70.0\\
NGC 7662 & 111.9 & 113.0 & 96.8 & 100.0 & 111.0 & 95.0\\
\hline
\end{tabular}
\tablenotetext{a}{\cite{sha85} and \cite{sha89}.}
\tablenotetext{b}{\cite{zha93}.}
\tablenotetext{c}{\cite{cor03}.}
\tablenotetext{d}{\cite{fre08}.}
\tablenotetext{e}{\cite{hen15}.}
\end{table*}
As alluded to in the introduction, our modeling methodology here is quite different than in Paper II. The main differences are the choice of SED, the calculation of the model emission lines, and the location of the gas boundaries. In Paper II, the SED was chosen to be a blackbody, whereas here the chosen SED (Rauch H-Ni or Rauch pg1159) more accurately represents the central star. With the code PANIC, each region's line-of-sight model emission line strengths were compared to the observations as opposed to the model emission line strengths as seen viewing the entire PN model. As in Paper II, the models are matter bounded. However, here we define inner and outer radii based on the measured distances observed from HST images. In Paper II, the inner radius was held fixed at 10$^{17}$ cm (0.0324 pc) and the outer radius was reached when the model emission line of [O~II] $\lambda$3727 matched the observation. These differences can very well explain why the final luminosities determined here for NGC 2440 and NGC 5315 differ significantly from those of Paper II.
\\\\Values of the final mass and radius as well as the main sequence lifespan, ZAMS mass, radius, luminosity, temperature, and spectral type can be estimated from the model-derived T and L values for each planetary nebula. Each of these derived properties is shown in Table~\ref{stellprop}. The final and ZAMS masses are estimated by interpolating between theoretical post-AGB evolutionary tracks on the H-R diagram shown in Figure~\ref{agbmods}. For NGC 5315, the mass estimates are from extrapolation since there are no post-AGB evolutionary tracks that have been successfully modeled with luminosities that high. Each track is from either \citet{vas94} (VW 1994, red solid lines), \citet{sch83} (S 1983, blue dash lines), or \cite{ber16} (MB 2016, green dash and purple dash-dot lines). Beneath each author on the plot is the ZAMS/final mass in solar masses following each evolutionary track from top to bottom. The ZAMS/final masses from different interpolations were averaged together when overlapping occurred since a single set of models was insufficient for all calculations. The final masses for four of our PNe match with the white dwarf mass peak, 0.565$M_{\astrosun}$, from \citet{lie05}. The current radius was calculated using the Stefan-Boltzmann law. Mass-luminosity and mass-radius relations from \citet{dem91} (appropriate for a given mass) were used to calculate the ZAMS luminosity and radius, respectively. The Stefan-Boltzmann law was also employed to calculate the ZAMS temperature. The main sequence lifespan was based on the approximate lifespan of the Sun, $\tau_{\astrosun}\approx$ M$_{\astrosun}$/L$_{\astrosun}$ = 9.5 Gyr, and the age of the Milky Way Galaxy, 13.2 Gyr, was used as a hard upper limit. Lastly, the ZAMS spectral type was determined from the ZAMS luminosity with the spectral type range within the brackets being based on the uncertainties in the luminosity.
\begin{table*}
\centering
\caption{Derived Stellar Parameters.}
\label{stellprop}
\begin{tabular}{ccccccc}
\hline
Stellar Property & IC 2165 & IC 3568 & NGC 2440 & NGC 5315 & NGC 5882 & NGC 7662\\
\hline
Final Mass (M$_{\astrosun}$)				& 0.56$^{+0.02}_{-0.01}$ & 0.55$^{+0.26}_{-0.01}$ & 0.66$^{+0.08}_{-0.06}$ & 1.17$^{+0.09}_{-0.09}$ & 0.56$^{+0.03}_{-0.01}$ & 0.62$^{+0.04}_{-0.06}$ \\
Current Radius (R$_{\astrosun}$) 				& 0.11$^{+0.08}_{-0.03}$ & 0.16$^{+0.36}_{-0.07}$ & 0.03$^{+0.01}_{-0.01}$ & 1.90$^{+0.53}_{-0.55}$ & 0.29$^{+0.15}_{-0.06}$ & 0.20$^{+0.07}_{-0.05}$ \\
Main Sequence Lifespan (Gyr) 				& 9.5$^{+3.7}_{-4.3}$ & 8.3$^{+4.9}_{-8.1}$ & 0.5$^{+0.8}_{-0.2}$ & 0.04$^{+0.01}_{-0.01}$ & 8.9$^{+3.5}_{-4.4}$ & 3.6$^{+1.9}_{-1.0}$ \\
ZAMS Mass (M$_{\astrosun}$)			& 0.99$^{+0.22}_{-0.15}$ & 1.04$^{+3.02}_{-0.23}$ & 2.81$^{+0.66}_{-0.81}$ & 6.55$^{+0.66}_{-0.65}$ & 1.02$^{+0.27}_{-0.11}$ & 1.81$^{+0.70}_{-0.42}$ \\
ZAMS Radius (R$_{\astrosun}$) 			& 1.05$^{+0.22}_{-0.15}$ & 1.10$^{+1.79}_{-0.23}$ & 2.36$^{+0.29}_{-0.40}$ & 3.78$^{+0.21}_{-0.21}$ & 1.08$^{+0.27}_{-0.11}$ & 1.85$^{+0.37}_{-0.25}$ \\
ZAMS $L_{star}$ ($L_{\astrosun}$) 		& 0.99$^{+1.21}_{-0.47}$ & 1.20$^{+247.43}_{-0.75}$ & 58.65$^{+75.32}_{-43.09}$ & 1618.68$^{+734.07}_{-546.60}$ & 1.09$^{+1.68}_{-0.39}$ & 10.52$^{+27.53}_{-6.81}$ \\
ZAMS Temperature (kK) 				& 5.6$^{+6.0}_{-4.0}$ & 5.8$^{+7.7}_{-0.7}$ & 10.4$^{+1.7}_{-2.2}$ & 18.9$^{+1.3}_{-1.3}$ & 5.7$^{+0.7}_{-0.3}$ & 7.6$^{+2.0}_{-1.3}$ \\
ZAMS Spectral Type 				& G2 [F6-G8] & G0 [B6-K3] & B8 [B7-A3] & B2 [B2-B3] & G1 [F5-G5] & A7 [B9-F3] \\
\hline
\end{tabular}
\end{table*}
\citet{amn95} suggested that an increase in progenitor mass would increase the probability of low mass companions, which would determine the axisymmetric morphology of the planetary nebula. As can be seen in Table~\ref{stellprop}, the central stars of IC 2165, IC 3568, and NGC 5882 are each solar-like while those of NGC 2440, NGC 5315, and NGC 7662 are super-solar. NGC 5882, NGC 2440, and NGC 5315 are the only PN in our sample to exhibit very asymmetric structures (Figure~\ref{fig:slits}), in support of the claim made by \citet{amn95}.
\\\\Given that the time it takes a high mass star to reach its maximum stellar temperature during the post-AGB phase is of order less than 500 years \cite[e.g.][]{ber16}, how likely is it that NGC 5315 is observed at such a low temperature? \citet{kas08} detected a ``hot bubble'' within NGC 5315 through the emission of X-rays. Such X-ray emitting regions are believed to arise from fast winds ($\sim$1000 km s$^{-1}$) during early PNe formation, shocking and sweeping up previously ejected, slower gas and dust, producing a bright rim \cite[e.g.][]{zhe96}. Using the expansion velocity from \citet{ack92} for the rim in NGC 5315, 40 km s$^{-1}$, and dividing the inner radius of the best fit full region model by this value gives an expansion time of about 300 years. This young age supports the possibility that the central star is still getting hotter. Additionally, since the central star of NGC 5315 is hydrogen deficient, the evolution of such stars have been shown to be slower than hydrogen rich stars \citep[e.g.][]{vas94}. Recent work by \citet{mad17} on measuring neutron-capture element abundances in NGC 5315 also couldn't rule out a high mass progenitor. However, the need for extrapolation of the evolutionary models, which show the evolution for hydrogen rich central stars, leaves the initial/final masses and properties derived from them questionable as to their accuracy.

\section{Summary and Conclusions}
We investigated the level of homogeneity of elements distributed in a small set of PN. We also modeled each PN to constrain the stellar luminosity and temperature. To test the homogeneity, the co-spatial observations described in Paper~I were divided into different spatial regions and spectra were extracted using an in-home Python script. The IRAF task \textit{splot} was used to measure the individual line strengths and these measurements served as input for the program ELSA, which calculated the nebular temperatures, densities, and abundances. Comparisons of the nebular properties among regions of each PN were then made. 
\\\\Next, we employed Cloudy to generate a model of each PN in order to determine the stellar properties using observational constraints. The code PANIC calculated an rms value that determined the effectiveness of each model to match each observation, and the rms value was used to choose the best parameters and estimate errors. Finally, three different density profiles (constant density, Gaussian density, and Gaussian with a power-law density profile) were used to see the effects each has on the best stellar parameters. 
\\\\The conclusions from our work are similar to Paper~III and are as follows. 
\begin{itemize}
\item The planetary nebulae in our sample are chemically homogeneous, which implies the shells of material that were ejected from their respective stars were well-mixed. This is in line with current theoretical work on the formation of planetary nebulae. The homogeneity also means that observations can be taken anywhere across the PN and the resulting abundance will accurately represent the nebula as a whole.
\item The constant density models can constrain the stellar parameters quite well depending on the planetary nebula. Also, the choice of density structure has only a small effect on the final stellar properties. This is advantageous since the model calculation time for non-constant density profiles is considerably higher than for constant density profiles. For NGC 2440 and IC 2165, the non-constant density profiles were disfavored given the much larger rms values compared to the constant density rms. This implies that they are better represented by the constant density profile even with the large difference in morphology.
\item The progenitor masses for the majority of our sample were around 1 M$_{\astrosun}$. This is expected since the observed distribution of white dwarf masses peaks at a progenitor mass around 1 M$_{\astrosun}$. The largest mass at 6.55 M$_{\astrosun}$ for NGC 5315 is somewhat questionable given the need to extrapolate from the post-AGB evolutionary models. This can only be improved with more high mass models being calculated.
\end{itemize}

\section*{Acknowledgements}

Support for Program number GO12600 was provided by NASA through a grant from the Space Telescope Science Institute, which is operated by the Association of Universities for Research in Astronomy, Incorporated, under NASA contract NAS5-26555. All authors are grateful to their home institutions for travel support, if provided. Most of the computing for this project was performed at the OU Supercomputing Center for Education and Research (OSCER) at the University of Oklahoma. Special thanks to the reviewer, Michael Richer, whose comments and suggestions improved this paper's quality.






\clearpage

\begin{deluxetable}{clccccccccccc}
\tabletypesize{\footnotesize}
\setlength{\tabcolsep}{0.07in}
\tablewidth{0in}
\tablecaption{\label{emisstab1all}}
\tablehead{
\multicolumn{13}{c}{Fluxes and Intensities}
}
\startdata
Wave & & & \multicolumn{2}{c}{IC2165 Full} & \multicolumn{2}{c}{IC2165 1} & \multicolumn{2}{c}{IC2165 2} & \multicolumn{2}{c}{IC2165 3} & \multicolumn{2}{c}{IC2165 4} \\
\cline{4-13}
(\AA) & ID & f($\lambda$) & F($\lambda$) & I($\lambda$) & F($\lambda$) & I($\lambda$) & F($\lambda$) & I($\lambda$) & F($\lambda$) & I($\lambda$) & F($\lambda$) & I($\lambda$) \\
\tableline
1485 & N IV] 	& 1.231 & 13.7 & 47.9$\pm$5.20 & 10.4 & 34.5$\pm$5.10 & 19.4 & 76.7$\pm$11.59 & 17.2 & 55.2$\pm$7.43 & 18.6 & 53.4$\pm$6.95 \\
1907 & C III] 	& 1.226 & 112 & 387$\pm$41 & 123 & 404$\pm$53 & 108 & 426$\pm$62 & 101 & 323$\pm$42 & 100 & 286$\pm$36 \\
1909 & C III] 	& 1.229 & 86.4 & 301$\pm$32 & 93.0 & 306$\pm$40 & 93.8 & 370$\pm$55 & 75.0 & 240$\pm$32 & 76.2 & 218$\pm$28 \\
3869 & [Ne III] 	& 0.224 & 59.4 & 76.8$\pm$5.40 & 65.8 & 84.1$\pm$6.29 & 48.3 & 64.0$\pm$7.00 & 47.7 & 60.5$\pm$7.18 & 44.6 & 55.3$\pm$6.24 \\
4363 & [O III] 	& 0.118 & 16.0 & 18.0$\pm$1.94 & 20.0 & 22.4$\pm$3.96 & 15.0 & 17.1$\pm$2.34 & 14.1 & 15.7$\pm$3.16 & 13.8 & 15.2$\pm$3.35 \\
4686 & He II 	& 0.036 & 60.0 & 62.2$\pm$2.97 & 58.4 & 60.5$\pm$3.59 & 80.5 & 83.8$\pm$4.75 & 68.3 & 70.7$\pm$4.14 & 78.7 & 81.2$\pm$3.70 \\
\enddata
\tablenotetext{a}{$\mathrm{L}$ogarithmic extinction at H$\beta$}
\tablenotetext{b}{Expected intrinsic H$\alpha$/H$\beta$ ratio at nebular temperature and density}
\tablenotetext{c}{ergs\ cm$^{-2}$ s$^{-1}$ in our extracted spectra}
\tablenotetext{d}{Blended 1907 and 1909}
\end{deluxetable}

\clearpage
\begin{deluxetable}{cccccccc}
\tabletypesize{\footnotesize}
\setlength{\tabcolsep}{0.07in}
\tablewidth{0in}
\tablecaption{\label{ionabun1all}}
\tablehead{\multicolumn{8}{c}{Ionic Abundances\tablenotemark{a}, Temperatures and Densities}
}
\startdata
Ion & IC2165 Full & IC2165 1 & IC2165 2 & IC2165 3 & IC2165 4 & IC2165 5 & IC2165 6\\ 
\tableline
He$^{+}$(5876) & 5.67$\pm$1.62(-2) & 5.12$\pm$1.50(-2) & 3.87$\pm$1.60(-2) & 3.95$\pm$1.30(-2) & 2.97$\pm$1.12(-2) & 6.86$\pm$1.07(-2) & 1.00$\pm$0.28(-1)\\
He$^{+2}$(4686) & 5.65$\pm$0.27(-2) & 5.47$\pm$0.33(-2) & 7.52$\pm$0.43(-2) & 6.42$\pm$0.39(-2) & 7.33$\pm$0.36(-2) & 3.58$\pm$0.28(-2) & \nodata\\
icf(He) & 1.00\tablenotemark{c} & 1.00\tablenotemark{c} & 1.00\tablenotemark{c} & 1.00\tablenotemark{c} & 1.00\tablenotemark{c} & 1.00\tablenotemark{c} & 1.00\tablenotemark{c}\\
O$^{+}$(3727) & 7.44$\pm$4.06(-6) & 7.91$\pm$3.45(-6) & 9.00$\pm$2.89(-6) & 5.80$\pm$3.51(-6) & 2.11$\pm$1.72(-6)  & 1.24$\pm$0.33(-5) & 1.48$\pm$0.77(-4)\\
O$^{+}$(7325) & 2.31$\pm$2.24(-5) & 8.55$\pm$4.08(-6) & 3.66$\pm$2.16(-6) & 9.65$\pm$8.20(-6) & 5.91$\pm$7.13(-6) & 2.02$\pm$0.79(-5) & 6.68$\pm$5.23(-5)\\
O$^{+}$(adopt) & 1.14$\pm$0.41(-5) & 8.06$\pm$2.76(-6) & 7.94$\pm$2.54(-6) & 6.90$\pm$3.60(-6) & 3.74$\pm$4.18(-6) & 1.44$\pm$0.33(-5) & 1.29$\pm$0.69(-4)\\ 
O$^{+2}$(5007) & 1.60$\pm$0.21(-4) & 1.71$\pm$0.37(-4) & 1.08$\pm$0.18(-4) & 1.31$\pm$0.32(-4) & 1.11$\pm$0.30(-4) & 2.13$\pm$0.41(-4) & 7.61$\pm$5.41(-4)\\
\enddata
\tablenotetext{a}{\footnotesize Abundances relative to H$^{+}$; n.nn$\pm$n.nn(-k) == (n.nn$\pm$n.nn) x 10$^{-k}$}
\tablenotetext{b}{\footnotesize Assumes the ICF(O) = 1}
\tablenotetext{c}{\footnotesize Assumed value}
\end{deluxetable}

\clearpage
\begin{deluxetable}{lcccccccc}
\tabletypesize{\footnotesize}
\setlength{\tabcolsep}{0.07in}
\tablewidth{0in}
\tablecaption{\label{tab:abun1all}}
\tablehead{
\multicolumn{9}{c}{Total Elemental Abundances}
}
\startdata
Parameter & IC2165 Full & IC2165 1 & IC2165 2 & IC2165 3 & IC2165 4 & IC2165 5 & IC2165 6 & Solar\tablenotemark{a}\\ 
\tableline
He/H (10$^{-2}$) 	& 11.30$^{+1.65}_{-1.65}$ & 10.60$^{+1.56}_{-1.56}$ & 11.40$^{+1.67}_{-1.67}$ & 10.40$^{+1.38}_{-1.38}$ & 10.30$^{+1.21}_{-1.21}$ & 10.40$^{+1.13}_{-1.13}$ & 10.00$^{+2.78}_{-2.78}$ & 8.51 \\
C/H (10$^{-4}$) 	& 4.14$^{+0.87}_{-0.87}$ & 3.09$^{+1.05}_{-1.05}$ & 3.81$^{+1.03}_{-1.03}$ & 3.83$^{+1.54}_{-1.54}$ & 2.91$^{+1.28}_{-1.28}$ & 4.23$^{+1.32}_{-1.32}$ & 89.00$^{+114.84}_{-114.84}$ & 2.69 \\
C/O 			& 1.221$^{+0.343}_{-0.343}$ & 0.842$^{+0.348}_{-0.348}$ & 1.117$^{+0.453}_{-0.453}$ & 1.052$^{+0.519}_{-0.519}$ & 0.735$^{+0.409}_{-0.409}$ & 1.223$^{+0.443}_{-0.443}$ & 10.253$^{+14.954}_{-14.954}$ & 0.550 \\
N/H (10$^{-4}$) 	& 1.18$^{+0.29}_{-0.29}$ & 0.53$^{+0.24}_{-0.24}$ & 1.21$^{+0.38}_{-0.38}$ & 1.49$^{+0.71}_{-0.71}$ & 1.05$^{+0.54}_{-0.54}$ & 0.67$^{+0.27}_{-0.27}$ & 2.44$^{+4.54}_{-4.54}$ & 0.68 \\
N/O 			& 0.348$^{+0.108}_{-0.108}$ & 0.144$^{+0.073}_{-0.073}$ & 0.353$^{+0.155}_{-0.155}$ & 0.409$^{+0.227}_{-0.227}$ & 0.264$^{+0.164}_{-0.164}$ & 0.192$^{+0.085}_{-0.085}$ & 0.281$^{+0.557}_{-0.557}$ & 0.138 \\
O/H (10$^{-4}$) 	& 3.39$^{+0.64}_{-0.64}$ & 3.67$^{+0.87}_{-0.87}$ & 3.41$^{+1.03}_{-1.03}$ & 3.64$^{+1.04}_{-1.04}$ & 3.96$^{+1.35}_{-1.35}$ & 3.46$^{+0.63}_{-0.63}$ & 8.68$^{+5.90}_{-5.90}$ & 4.90 \\
Ne/H (10$^{-5}$) 	& 5.36$^{+1.26}_{-1.26}$ & 5.61$^{+1.49}_{-1.49}$ & 5.63$^{+1.84}_{-1.84}$ & 5.41$^{+1.78}_{-1.78}$ & 5.86$^{+2.22}_{-2.22}$ & 6.03$^{+1.31}_{-1.31}$ & 19.50$^{+16.00}_{-16.00}$ & 8.51 \\
\enddata
\tablenotetext{a}{\citet{asp09}}
\tablenotetext{b}{\footnotesize Assumes the ICF(O) = 1}
\end{deluxetable}
\clearpage

\begin{deluxetable}{clccccc}
\tabletypesize{\footnotesize}
\setlength{\tabcolsep}{0.07in}
\tablewidth{0in}
\tablecaption{\label{modline1all}}
\tablehead{
\multicolumn{7}{c}{Constant Density Model Emission Lines Compared to Observations}
}
\startdata
Wave & & IC2165 Full & IC2165 1 & IC2165 2 & IC2165 3 & IC2165 4 \\ 
\cline{3-7}
(\AA) & ID & Model/Observed & Model/Observed & Model/Observed & Model/Observed & Model/Observed \\
\tableline
1485 & N IV] 				& 1.00$\pm$0.11 & 1.00$\pm$0.15 				& 1.00$\pm$0.15 			& 1.00$\pm$0.13 			& 1.01$\pm$0.13 \\     
1907 & [C III]					& 1.01$\pm$0.11 & 0.96$\pm$0.13 				& 1.05$\pm$0.15 			& 1.03$\pm$0.13 			& 1.06$\pm$0.13 \\    
1909\tablenotemark{b} & C III]  		& 1.00$\pm$0.11 & 0.97$\pm$0.13 				& 1.04$\pm$0.15 			& 1.03$\pm$0.13 			& 1.06$\pm$0.13 \\        
3869 & [Ne III]  				& 1.00$\pm$0.07 & 1.00$\pm$0.07 				& 1.00$\pm$0.11 			& 1.00$\pm$0.12 			& 1.01$\pm$0.11 \\  
4363\tablenotemark{b} & [O III]  	& 1.01$\pm$0.11 & 0.87$\pm$0.18	 			& 1.11$\pm$0.14			& 1.04$\pm$0.20 			& 1.10$\pm$0.22 \\       
4686 & He II  				& 1.00$\pm$0.05 & 1.09$\pm$0.06\tablenotemark{a} 	& 0.81$\pm$0.06\tablenotemark{a} & 0.97$\pm$0.06			& 0.85$\pm$0.05\tablenotemark{a} \\
\enddata
\tablenotetext{a}{Modeled emission line intensity outside observed error bar.}
\tablenotetext{b}{This line was only included in a diagnostic for the rms calculation for reasons discussed in the text.}
\tablenotetext{c}{Blended 1907 and 1909.}
\end{deluxetable}
\clearpage

\begin{deluxetable}{cccccc}
\tabletypesize{\footnotesize}
\tablewidth{0in}
\tablecaption{\label{condenmods1all}}
\tablehead{
\multicolumn{6}{c}{Constant Density Models}
}
\startdata
Parameter & IC2165 Full & IC2165 1 & IC2165 2 & IC2165 3 & IC2165 4 \\
\tableline
$T_{star}$ (kK) & 110.0$^{+7.0}_{-17.0}$ & 110.0 & 110.0 & 110.0& 110.0  \\
$L_{star}$ (log[L/$L_{\astrosun}$]) & 3.17$^{+0.20}_{-0.14}$ & 3.17 & 3.17 & 3.17 & 3.17  \\
Distance (kpc) & 2.39$^{+0.36}_{-0.33}$ & 2.39 & 2.39 & 2.39 & 2.39 \\
$H_{den}$ (log[$H_{density}$]) & 3.79$^{+0.02}_{-0.03}$ & 3.75$^{+0.04}_{-0.04}$ & 4.06$^{+0.05}_{-0.07}$ & 3.70$^{+0.04}_{-0.04}$ & 3.75$^{+0.04}_{-0.05}$ \\
Inner Radius ($10^{-2}$pc) & 2.60$^{+0.06}_{-0.06}$ & 2.56$^{+0.11}_{-0.15}$ & 2.60$^{+0.05}_{-0.04}$ & 2.38$^{+0.15}_{-0.16}$ & 2.47$^{+0.13}_{-0.13}$ \\
Outer Radius ($10^{-2}$pc) & 3.09$^{+0.05}_{-0.05}$ & 3.09$^{+0.08}_{-0.07}$ & 2.73$^{+0.04}_{-0.05}$ & 3.11$^{+0.12}_{-0.11}$ & 2.96$^{+0.12}_{-0.12}$ \\
Filling Factor ($10^{-1}$) & 9.89$^{+0.11}_{-0.99}$ & 10.00$^{+0.00}_{-0.16}$ & 9.49$^{+0.51}_{-2.97}$ & 10.00$^{+0.00}_{-0.17}$ & 10.0$^{+0.00}_{-0.25}$ \\
\enddata
\tablenotetext{a}{Model Returned High Density Limit Default: 20000 cm$^{-3}$.}
\tablenotetext{DMS}{Using the formalism of \citet{del14}.}
\tablenotetext{b}{\footnotesize Error unknown since observed value is assumed.}
\tablenotetext{c}{\footnotesize Assumes the ICF(O) = 1}
\end{deluxetable}
{.\pagebreak}







\bsp	
\label{lastpage}
\end{document}